\documentclass[twocolumn,prl,superscriptaddress,longbibliography]{revtex4-2}
\usepackage[colorlinks=true, citecolor=blue, urlcolor=blue, linkcolor=red]{hyperref}
\usepackage{graphicx,amsmath,amssymb,orcidlink}
\renewcommand{\section}[1]{{\par\it #1.---}\ignorespaces}

\begin{document}
\title{Non-Canonical Quantum Otto Engine}
\author{Wei Wu\orcidlink{0000-0002-5989-9458}}
\affiliation{Key Laboratory of Quantum Theory and Applications of Ministry of Education, Lanzhou Center for Theoretical Physics, Gansu Provincial Research Center for Basic Disciplines of Quantum Physics, Key Laboratory of Theoretical Physics of Gansu Province, Lanzhou University, Lanzhou 730000, China}
\author{Jun-Hong An\orcidlink{0000-0002-3475-0729}}
\affiliation{Key Laboratory of Quantum Theory and Applications of Ministry of Education, Lanzhou Center for Theoretical Physics, Gansu Provincial Research Center for Basic Disciplines of Quantum Physics, Key Laboratory of Theoretical Physics of Gansu Province, Lanzhou University, Lanzhou 730000, China}

\begin{abstract}
The fast-developing quantum technology is propelling thermal machine, which lies at the heart of thermodynamics, into a new golden era. Being stochastic because of the non-negligible thermal and quantum fluctuations, the efficiency and power of a finite-time quantum heat engine exhibit a trade-off, preventing the simultaneous achievement of high power and high efficiency. Conventional studies treat the statistics of a heat engine by assuming canonical thermalization, which is valid only under weak-coupling conditions. We here investigate the noncanonical effect on the statistics of the efficiency and power for a finite-time quantum Otto cycle. It is revealed that their average values and variances are highly controllable by engineering the energy spectrum structure of the composite system consisting of the working substance and the bath. Near the quantum critical point induced by the formation of a bound state in the energy spectrum, the average efficiency and power can be boosted simultaneously, meanwhile, both of their variances are decreased. Deepening our understanding of the usefulness of quantum reservoir engineering in thermodynamics, our result lays the foundation for the realization of efficient and stable quantum energy devices.
\end{abstract}
\maketitle

\section{Introduction}\label{sec:sec1}
Since the first industrial revolution, exploring the properties and performances of heat engines, which are thermodynamical devices that convert energy from one form into another, have become fundamentally essential in the research field of thermodynamics~\cite{CANGEMI20241}. In recent years, the rapid development of various nanoscale technologies has allowed devices to acquire the advantages induced by quantum characteristics~\cite{doi:10.1126/science.1078955,PhysRevLett.112.030602,Campisi2016,Dillenschneider_2009,PhysRevE.93.052103,PhysRevE.96.022143,PhysRevLett.104.207701,GashuFeyisa_2024,PhysRevResearch.5.043185,Myers_2022,PhysRevLett.134.010407,PhysRevLett.130.240401,PhysRevLett.130.110402}, which do not have classical counterparts and beat the traditional bound. The exploration of these quantum characteristics breeds a new quantum industrial revolution, which paves the way to a higher efficiency for the conversion of heat into work and to explore fundamental aspects of thermodynamics in the quantum regime.

The efficiency of a traditional macroscopic heat machine is deterministic. By contrast, for a heat engine operating at the level of microscopic scales, its work and heat commonly become stochastic due to the presence of thermal and quantum fluctuations~\cite{Seifert_2012,RevModPhys.81.1665,RevModPhys.83.771}. Thus, it is more scientific to treat efficiency and power as stochastic quantities~\cite{Verley2014,Martínez2016,PhysRevResearch.2.032062,Denzler_2021,PhysRevA.105.022609,Denzler_2024}. In most of the previous studies of quantum heat engines, efficiency is defined as the ratio between the average extracted work and the average heat absorbed from the hot bath~\cite{PhysRevE.76.031105,10.1063/5.0192075,Wiedmann_2020,PhysRevA.99.052106,HamedaniRaja_2021,PhysRevA.109.022207,PhysRevE.95.032139}. However, these average values cannot provide a full description of stochastic properties. In order to accurately assess the performance of a quantum heat engine, investigation of the statistical features of efficiency and power is necessary~\cite{PhysRevResearch.2.032062,Denzler_2021,PhysRevA.105.022609,Denzler_2024}. The fluctuation characters of an Otto heat engine have been studied~\cite{PhysRevResearch.2.032062,Denzler_2021,PhysRevA.105.022609,Denzler_2024}. All of these studies have employed the canonical thermalization treatment, which assumes that the interaction between the working substance and the bath is so weak that the Born-Markov approximation can be safely adopted. However, as demonstrated in Refs.~\cite{PhysRevE.90.022122,PhysRevA.89.012128,PhysRevResearch.4.023141}, such a canonical thermalization approximation generally breaks down in the strong-coupling regime in which the non-Markovianity induced noncanonical thermalization occurs. An interesting question naturally arises: what is the influence of the noncanonical effect on the performance of a quantum heat engine.

\begin{figure}[tbp]
\centering
\includegraphics[angle=0,width=0.475\textwidth]{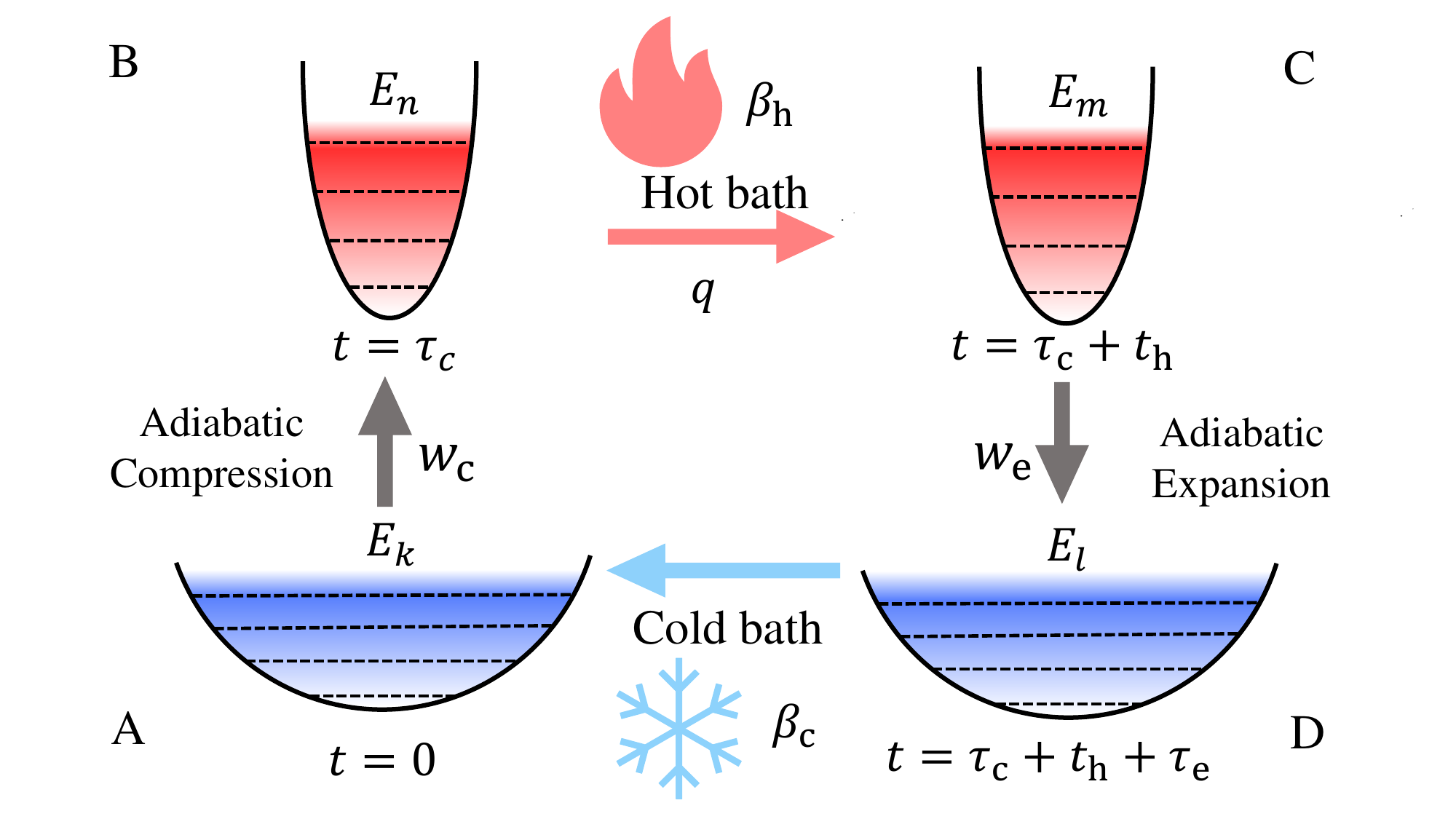}
\caption{Diagrammatic sketch of the quantum Otto cycle.}\label{fig:fig1}
\end{figure}

In this Letter, going beyond the usual limitation of canonical thermalization treatment, we investigate the statistics of efficiency and power for a finite-time quantum Otto heat engine in the non-Markovian dynamical regime. We find that they depend sensitively on the energy-spectrum feature of the total system consisting of the working substance and the hot bath. With the formation of a bound state in the energy spectrum, which leads to a quantum phase transition~\cite{Rançon_2013,PhysRevB.84.174301,PhysRevApplied.17.034073,PhysRevA.109.042202}, the average efficiency and the average power exhibit a cusp structure, meanwhile, their variances can be accordingly reduced. This offers us a method of harvesting a highly stable quantum heat engine with simultaneously enhanced efficiency and power, which is hard to achieve in conventional quantum heat engines \cite{Krishnamurthy2023,vlb3-p63l,pw37-jjdp}, by engineering the energy-spectrum feature of the total system via the well developed technique of quantum reservoir engineering~\cite{ER1,Kienzler53,PhysRevA.78.010101}.

\section{Finite-time quantum Otto engine}\label{sec:sec2}
We consider a finite-time quantum Otto engine, in which the working substance is a quantum harmonic oscillator (see Fig.~\ref{fig:fig1}). Its Hamiltonian is $\hat{H}_{\text{s}}(t)=\frac{1}{2}[\hat{p}^{2}+\omega(t)^{2}\hat{x}^{2}]$ with $\hbar=m=1$. $\hat{x}$ and $\hat{p}$ are, respectively, the position and momentum operators satisfying the canonical commutation relation $[\hat{x},\hat{p}]=i$. $\omega(t)$ is a time-dependent frequency and serves as a controllable parameter to extract work from the working substance. The quantum Otto engine has four strokes. The working substance is initially in a Gibbs state at the same temperature as that of a cold bath, i.e., $T_{\text{c}}=1/\beta_{\text{c}}$ with $k_{\text{B}}=1$. In the first stroke from $t=0$ to $\tau_\text{c}$, it undergoes a finite-time compression process, in which the frequency increases from $\omega_{\text{i}}$ to $\omega_{\text{f}}$. Being isolated from the bath, its dynamics is unitarily governed by $\hat{H}_{\text{s}}(t)$~\cite{PhysRevE.77.021128,DEFFNER2010200}. In the second stroke from $t=\tau_\text{c}$ to $\tau_\text{c}+t_\text{h}$, the working substance in a fixed frequency $\omega_{\text{f}}$ interacts with a hot bath at a temperature $T_{\text{h}}=1/\beta_{\text{h}}$ and absorbs heat from the bath. In the third stroke from $t=\tau_\text{c}+t_\text{h}$ to $\tau_\text{c}+t_\text{h}+\tau_\text{e}$, it is decoupled from the bath and undergoes a finite-time expansion process, in which the frequency decreases from $\omega_{\text{f}}$ back to $\omega_{\text{i}}$. This stroke is an inverse process of the first stroke. In the fourth stroke from $t=\tau_\text{c}+t_\text{h}+\tau_\text{e}$ to $\tau_\text{c}+t_\text{h}+\tau_\text{e}+t_\text{c}$, the working substance in a fixed frequency $\omega_{\text{i}}$ interacts with the cold bath. We assume that their interaction is very weak and the decoherence time is so short that the working substance experiences a canonical thermalization~\cite{PhysRevE.90.022122,PhysRevResearch.4.023141,PhysRevLett.133.050401}. This is necessary to make the working substance at the end of the fourth stroke return to its initial state, which ensures the closure of the cycle.

The total Hamiltonian of the working substance and the hot bath in the second stroke is
\begin{equation}\label{eq:eq1}
\hat{H}=\omega_{\text{f}}\hat{a}^{\dagger}\hat{a}+\sum_{k}\omega_{k}\hat{b}_{k}^{\dagger}\hat{b}_{k}+\sum_{k}(g_{k}\hat{a}^{\dagger}\hat{b}_{k}+\text{H}.\text{c}.),
\end{equation}
where $\hat{a}=(\omega_{\text{f}}\hat{x}+i\hat{p})/\sqrt{2\omega_{\text{f}}}$ and $\hat{b}_{k}$ are the annihilation operators of the working substance and the $k$th mode with frequency $\omega_{k}$ of the hot bath, respectively, and $g_{k}$ is their coupling strength. After tracing the degrees of freedom of the bath, the  exact dynamics of the working substance is governed by the master equation~\cite{PhysRevLett.133.050401,PhysRevA.76.042127}
\begin{equation}\label{eq:eq2}
\dot{\rho}(t)=-i\Omega(t)[\hat{a}^{\dagger}\hat{a},\rho(t)]+\{[\gamma(t)+\tfrac{\gamma_{\beta}(t)}{2}]\check{\mathcal{L}}_{\hat{a}}+\tfrac{\gamma_{\beta}(t)}{2}\check{\mathcal{L}}_{\hat{a}^\dag}\}\rho(t),
\end{equation}
where $\check{\mathcal{L}}_{\hat{o}}\cdot=2\hat{o}\cdot\hat{o}^\dag-\{\hat{o}^\dag\hat{o},\cdot\}$ is the Lindblad superoperator, $\Omega(t)=-\text{Im}[\dot{u}(t)/u(t)]$ is the renormalized frequency, and $\gamma(t)=-\text{Re}[\dot{u}(t)/u(t)]$ and $\gamma_{\beta}(t)=\dot{v}(t)+2v(t)\gamma(t)$ are the bath-induced dissipation and noise coefficients, respectively. The functions $u(t)$ and $v(t)$ satisfy
\begin{eqnarray}
\dot{u}(t)+i\omega_{\text{f}}u(t)+\int_{0}^{t}d\tau\mu(t-\tau)u(\tau)=0,\label{eq:eq3}\\
v(t)=\int_{0}^{t}dt_{1}\int_{0}^{t}dt_{2}u^{*}(t_{1})\nu(t_{1}-t_{2})u(t_{2}),\label{eq:eq4}
\end{eqnarray}
with $u(0)=1$, $\mu(x)\equiv\int_{0}^{\infty}d\omega \mathcal{J}(\omega)e^{-i\omega x}$, and $\nu(x)\equiv\int_{0}^{\infty}d\omega \mathcal{J}(\omega)e^{-i\omega x}/(e^{\beta_{\text{h}}\omega}-1)$. Here, $\mathcal{J}(\omega)\equiv\sum_{k}|g_{k}|^{2}\delta(\omega-\omega_{k})$ is the spectral density.  We consider it to have an Ohmic-family form $\mathcal{J}(\omega)=\alpha\omega^{s}\omega_{c}^{1-s}e^{-\omega/\omega_{c}}$, where $\alpha$ is a dimensionless coupling constant, $\omega_{c}$ is a cutoff frequency, and $s$ is an Ohmicity index. The convolution in Eqs. \eqref{eq:eq3} and \eqref{eq:eq4} renders the dynamics non-Markovian. Determined by Eq. \eqref{eq:eq4}, the working substance equilibrates to a steady state at the end of the second stroke. A method widely used to treat this equilibration is the Born-Markov approximation, under which the steady state is uniquely a Gibbs state at temperature $T_\text{h}$ \cite{PhysRevResearch.2.032062,Denzler_2021,PhysRevA.105.022609,Denzler_2024}. Here, we go beyond this approximation and investigate the strong-coupling effect on the performance of the finite-time quantum Otto engine.

\section{Efficiency and power statistics}\label{sec:sec3}
The quantum fluctuation makes heat and work stochastic, leading to the randomness of the efficiency and power of quantum engines. We use the two-point measurement scheme~\cite{PhysRevResearch.2.032062,Denzler_2021,PhysRevA.105.022609,Denzler_2024,PhysRevLett.92.230602,PhysRevE.75.050102} to derive the joint distribution function for the efficiency and power of the quantum Otto engine. Suppose that the results obtained from the projective measurements of energy of the working substance at time $t=0$, $\tau_{\text{c}}$, $\tau_{\text{c}}+t_{\text{h}}$, and $\tau_{\text{c}}+t_{\text{h}}+\tau_{\text{e}}$ are $E_k$, $E_n$, $E_m$, and $E_l$, respectively, where $E_{j}$ satisfy $\hat{H}_{\text{s}}|E_{j}\rangle=E_{j}|E_{j}\rangle$, with $j=\{k,n,m,l\}$. The stochastic works in the compression and expansion processes are defined as $w_{\text{c}}=E_{n}-E_{k}$ and $w_{\text{e}}=E_{l}-E_{m}$. The stochastic heat absorbed from the hot bath is defined as $q=E_{m}-E_{n}$. The joint probability distribution $P(W,Q)$ of the total stochastic work and the stochastic heat is
\begin{equation}\label{eq:eq5}
\begin{split}
P(W,Q)=&\sum_{k,n,m,l}p_{k}(\beta_{\text{c}},\omega_{\text{i}})p_{kn}(\tau_{\text{c}})p_{nm}(\beta_{\text{h}},\omega_{\text{f}})\\
&\times p_{ml}(\tau_{\text{e}})\delta(W-w_{\text{e}}-w_{\text{c}})\delta(Q-q),
\end{split}
\end{equation}
where $p_{k}(\beta_{\text{c}},\omega_{\text{i}})=\frac{\bar{n}(\beta_{\text{c}},\omega_{\text{i}})^{k}}{[\bar{n}(\beta_{\text{c}},\omega_{\text{i}})+1]^{k+1}}$ with $\bar{n}(\beta_{\text{c}},\omega_{\text{i}})=(e^{\beta_{\text{c}}\omega_{\text{i}}}-1)^{-1}$ is the probability of obtaining $E_{k}$ at $t=0$ and $p_{jj'}$ is the conditional probability of obtaining $E_{j'}$ when $E_j$ is obtained in the previous-step measurement. $p_{jj'}(\tau_{\ell})$ with $\ell=\{\text{c},\text{e}\}$ can be expressed by the so-called generating function $\mathcal{G}_{\ell}(a,b)=\sum_{m,m'}p_{mm'}(\tau_{\ell})a^{m}b^{m'}$ as $p_{jj'}(\tau_\ell)=\tfrac{\partial^2\mathcal{G}_\ell(a,b)}{\partial a^j\partial b^{j'}}\big|_{a=b=1}$. It has been derived that $\mathcal{G}^{2}_{\ell}(a,b)=2[\mathcal{Q}_{\ell}(1-a^{2})(1-b^{2})+(1+a^{2})(1+b^{2})-4ab]^{-1}$, where $\mathcal{Q}_{\ell}=\{\omega^{2}(t_0)[\omega^{2}(\tau_{\ell})X^{2}(\tau_{\ell})+\dot{X}^{2}(\tau_{\ell})]+[\omega^{2}(\tau_{\ell})Y^{2}(\tau_{\ell})+\dot{Y}^{2}(\tau_{\ell})]\}/[2\omega(t_0)\omega(\tau_{\ell})]$ with $X(t)$ and $Y(t)$ being the solutions of $\ddot{c}(t)+\omega^{2}(t)c(t)=0$ under the conditions $X(t_0)=\dot{Y}(t_0)=0$ and $\dot{X}(t_0)=Y(t_0)=1$~\cite{PhysRevE.77.021128,SupplementalMaterial}. We explicitly take $\omega^2(t)=\omega^{2}(t_0)-[\omega^{2}(t_0)-\omega^{2}(\tau_{\ell})]t/\tau_\ell$ with $\omega(t_{0})=\omega_{\text{i}}$, $\omega(\tau_{\text{c}})=\omega_{\text{f}}$ for the first stroke and $\omega(t_{0})=\omega_{\text{f}}$, $\omega(\tau_{\text{e}})=\omega_{\text{i}}$ for the third stroke. Under such a linear driving scheme, $X(t)$ and $Y(t)$ are analytically solvable. It is necessary to stress that our result is generalizable to other driving schemes. Measuring the degree of adiabaticity for the process of work done, $\mathcal{Q}_{l}$ has $\lim_{\tau_{\ell}\rightarrow\infty}\mathcal{Q}_{\ell}\rightarrow 1$ in the adiabatic limit. To investigate the strong coupling effect in the second stroke on the engine performance, we exactly derive the steady-state solution of Eq. \eqref{eq:eq2} as
\begin{equation}\label{eq:eqpnm}
\begin{split}
p_{nm}&(\beta_{\text{h}},\omega_{\text{f}})=M(\infty)J_{2}(\infty)^{m}[1-J_{3}(\infty)]^{n}\\
&\times\sum_{q=0}^{\min\{n,m\}}C_{q}^{n}C_{q}^{m}\bigg{\{}\frac{|J_{1}(\infty)|^{2}}{[1-J_{3}(\infty)]J_{2}(\infty)}\bigg{\}}^{q},
\end{split}
\end{equation}
where $M(t)=[1+v(t)]^{-1}$, $J_{1}(t)=M(t)u(t)$, $J_{2}(t)=M(t)v(t)$, and $J_{3}(t)=M(t)|u(t)|^{2}$. Obviously, $p_{nm}(\beta_{\text{h}},\omega_{\text{f}})$ is sensitively dependent on the initial state $|E_n\rangle$, which means the breakdown of the usually assumed canonical thermalization. Such a noncanonical equilibration is purely induced by the non-Markovianity and is not predictable by the Born-Markov approximate treatment.  

To facilitate the study of efficiency and power statistics, we define the characteristic function of $P(W,Q)$ as $\chi(\xi,\zeta)=\int dW\int dQP(W,Q)e^{-\xi W-\zeta Q}$. Then the stochastic efficiency and power are~\cite{PhysRevA.105.022609}
\begin{equation}\label{eq:eq6}
\eta=-W/\langle Q\rangle,~~~~\mathcal{P}=-W/\tau_{\text{tot}},
\end{equation}
where $\langle Q\rangle=-{\partial\ln\chi(\xi,\zeta)\over \partial\zeta}\big|_{\xi=\zeta=0}$ is the average heat and $\tau_{\text{tot}}=\tau_{\text{c}}+t_{\text{h}}+\tau_{\text{e}}+t_{\text{c}}$ is the total cycle time. The efficiency and power of the quantum Otto engine are denoted by their averages as $\langle\eta\rangle$ and $\langle \mathcal{P}\rangle$. Their fluctuations are $\epsilon_{\eta}^{2}=\delta^{2}\eta/\langle\eta\rangle^{2}$ and $\epsilon_{\mathcal{P}}^{2}=\delta^{2}\mathcal{P}/\langle\mathcal{P}\rangle^{2}$~\cite{PhysRevResearch.3.L032041,PhysRevLett.121.120601,PhysRevE.103.032130}, where $\delta^{2}\theta\equiv\langle\theta^{2}\rangle-\langle\theta\rangle^{2}$. It is easy to prove that $\epsilon_{\eta}=\epsilon_{\mathcal{P}}\equiv\epsilon$, with
\begin{equation}\label{eq:eq7}
\epsilon=(\langle W^{2}\rangle-\langle W\rangle^{2})^{1/2}/\langle W\rangle,
\end{equation}
which is the relative fluctuation of the work and reflects the stability of the heat engine~\cite{PhysRevLett.121.120601,PhysRevResearch.3.L032041}. Our goal is to realize an Otto engine that operates with a high efficiency $\langle \eta\rangle$, a high power $\langle \mathcal{P}\rangle$, and a low fluctuation $\epsilon$. 

\section{Trade-off relation} The mean entropy production $\sigma=(\beta_{\text{c}}-\beta_{\text{h}})\langle Q\rangle+\beta_{\text{c}}\langle W\rangle$ exerts a lower bound on $\epsilon$ as $\epsilon^{2}\geq{2}/{\sigma}$~\cite{PhysRevLett.114.158101,PhysRevE.93.052145,PhysRevE.103.022136,PhysRevLett.123.090604}. This is the thermodynamic uncertainty relation for fluctuation, characterizing the minimal price of the entropy production one needs to pay to reduce the fluctuation of the heat engine. Noticing $\langle Q\rangle\geq0$ for heat engine and using $\sigma\geq0$, one easily proves that $\langle \eta\rangle\leq\eta_{\text{C}}$, with $\eta_{\text{C}}\equiv1-\beta_{\text{h}}/\beta_{\text{c}}$ being the Carnot efficiency~\cite{Sinitsyn_2011,Campisi_2014}. Being consistent with the second law of thermodynamics, the thermodynamic uncertainty relation exerts a trade-off for $\langle \eta\rangle$, $\langle \mathcal{P}\rangle$, and $\epsilon$ as
\begin{equation}\label{eq:eq12}
\epsilon^{2}\geq\epsilon_{\text{min}}^{2}=\frac{2}{\beta_{\text{c}}\tau_{\text{tot}}\langle \mathcal{P}\rangle}\frac{\langle \eta\rangle}{\eta_{\text{C}}-\langle \eta\rangle},
\end{equation}
A similar trade-off relation has been previously reported in classical heat engines~\cite{PhysRevLett.121.120601,PhysRevLett.120.190602}. Here, we generalize this relation to the quantum realm. Equation~(\ref{eq:eq12}) implies that it is not possible to construct a heat engine whose average efficiency is equal to the Carnot efficiency with finite fluctuation $\epsilon$. Considering the times $t_{\text{h},\text{c}}$ are much shorter than the compression and expansion times $\tau_{\text{c},\text{e}}$~\cite{PhysRevE.98.032121,PhysRevE.99.022110,PhysRevE.100.062140,PhysRevE.100.032144}, we use the approximation $\tau_{\text{tot}}\simeq\tau_{\text{c}}+\tau_{\text{e}}$ and $\tau_{\text{c}}=\tau_{\text{e}}\equiv\tau$.

\section{Markovian limit}\label{sec:sec4}
When the coupling between the working substance and the hot bath is weak and the correlation time scale of the hot bath is smaller than that of the working substance, we safely apply the Born-Markov approximation to $u(t)$ and obtain $u_\text{BM}(t)= \exp\{-\kappa t-i[\omega_{\text{f}}+\Delta(\omega_{\text{f}})]t\}$ with $\kappa=\pi \mathcal{J}(\omega_\text{f})$ being the decay rate and $\Delta(\omega_{\text{f}})=\mathcal{P}\int _0^\infty {\mathcal{J}(\omega)\over \omega_{\text{f}}-\omega}d\omega$ being the frequency shift~\cite{PhysRevE.90.022122}. It leads to $v(\infty)=\bar{n}(\beta_{\text{h}},\omega_{\text{f}})$ and the canonical thermalization of the working substance in the second stroke. Thus, we obtain $p_{nm}^{\text{BM}}(\beta_{\text{h}},\omega_{\text{f}})=\frac{\bar{n}(\beta_{\text{h}},\omega_{\text{f}})^{m}}{[\bar{n}(\beta_{\text{h}},\omega_{\text{f}})+1]^{m+1}}$ regardless of the initial state $|E_n\rangle$, which is consistent with the previous results~\cite{PhysRevResearch.2.032062,Denzler_2021,PhysRevA.105.022609,Denzler_2024,PhysRevE.103.032130}. It leads to~\cite{SupplementalMaterial}
\begin{equation}\label{eq:eq8}
\chi_\text{BM}(\xi,\zeta)=\frac{(e^{\beta_{\text{c}}\omega_{\text{i}}}-1)(e^{\beta_{\text{h}}\omega_{\text{f}}}-1)}{e^{\beta_{\text{c}}\omega_{\text{i}}+\beta_{\text{h}}\omega_{\text{f}}}}\mathcal{G}_{\text{c}}(a_{\text{c}},b_{\text{c}})\mathcal{G}_{\text{e}}(a_{\text{e}},b_{\text{e}}),
\end{equation}
where $a_{\text{c}}=e^{-\omega_{\text{i}}(\beta_{\text{c}}-\xi)}$, $b_{\text{c}}=e^{-\omega_{\text{f}}(\xi-\zeta)}$, $a_{\text{e}}=e^{-\omega_{\text{f}}(\beta_{\text{h}}-\xi+\zeta)}$, and $b_{\text{e}}=e^{-\omega_{\text{i}}\xi}$. Then it is straightforward to derive 
$\langle W_\text{BM}\rangle=\frac{1}{2}(\mathcal{Q}_{\text{c}}\omega_{\text{f}}-\omega_{\text{i}})\coth\frac{\beta_{\text{c}}\omega_{\text{i}}}{2}+\frac{1}{2}(\mathcal{Q}_{\text{e}}\omega_{\text{i}}-\omega_{\text{f}})\coth\frac{\beta_{\text{h}}\omega_{\text{f}}}{2}$, $\langle Q_\text{BM}\rangle=\frac{\omega_{\text{f}}}{2}(\coth\frac{\beta_{\text{h}}\omega_{\text{f}}}{2}-\mathcal{Q}_{\text{c}}\coth\frac{\beta_{\text{c}}\omega_{\text{i}}}{2})$, and
\begin{equation}\label{eq:eqvar}
\begin{split}
&\delta^{2}W_\text{BM}=\frac{2e^{\beta_{\text{h}}\omega_{\text{f}}+\beta_{\text{c}}\omega_{\text{i}}}}{(e^{\beta_{\text{h}}\omega_{\text{f}}}-1)^{2}(e^{\beta_{\text{c}}\omega_{\text{i}}}-1)^{2}} \Big{\{}2(\mathcal{Q}_{\text{c}}+\mathcal{Q}_{\text{e}})\\
&~~~~~~\times \omega_{\text{i}}\omega_{\text{f}}-(\mathcal{Q}^{2}_{\text{c}}+1)\omega^{2}_{\text{f}}-(\mathcal{Q}^{2}_{\text{e}}+1)\omega^{2}_{\text{i}}\\
&~~~~~~+\cosh(\beta_{\text{c}}\omega_{\text{i}})[\mathcal{Q}_{\text{e}}^{2}\omega_{\text{i}}^{2}-2\mathcal{Q}_{\text{e}}\omega_{\text{i}}\omega_{\text{f}}-(\mathcal{Q}_{\text{c}}^{2}-2)\omega^{2}_{\text{f}}]\\
&~~~~~~+\cosh(\beta_{\text{h}}\omega_{\text{f}})\Big{(}\mathcal{Q}_{\text{c}}^{2}\omega_{\text{f}}^{2}-2\mathcal{Q}_{\text{c}}\omega_{\text{i}}\omega_{\text{f}}-(\mathcal{Q}_{\text{e}}^{2}-2)\omega_{\text{i}}^{2}\\
&~~~~~~+\cosh(\beta_{\text{c}}\omega_{\text{i}})\big{[}(\mathcal{Q}_{\text{e}}^{2}-1)\omega_{\text{i}}^{2}+(\mathcal{Q}_{\text{c}}^{2}-1)\omega_{\text{f}}^{2}\big{]}\Big{)}\Big{\}}.
\end{split}
\end{equation}
It is easy to verify that $\langle\eta_\text{BM}\rangle=1-\omega_\text{i}/\omega_\text{f}$, $\langle\mathcal{P}_\text{BM}\rangle=0$, and $\epsilon^{2}_\text{BM}=[\sinh^{-2}(\frac{\beta_{\text{c}}\omega_{\text{i}}}{2})+\sinh^{-2}(\frac{\beta_{\text{h}}\omega_{\text{f}}}{2})]/[\coth(\frac{\beta_{\text{c}}\omega_{\text{i}}}{2})-\coth(\frac{\beta_{\text{h}}\omega_{\text{f}}}{2})]^{2}$ in the adiabatic limit $\lim_{\tau\rightarrow\infty}\mathcal{Q}_\text{c/e}=1$~\cite{PhysRevLett.109.203006}.

\begin{figure}
\centering
\includegraphics[angle=0,width=0.475\textwidth]{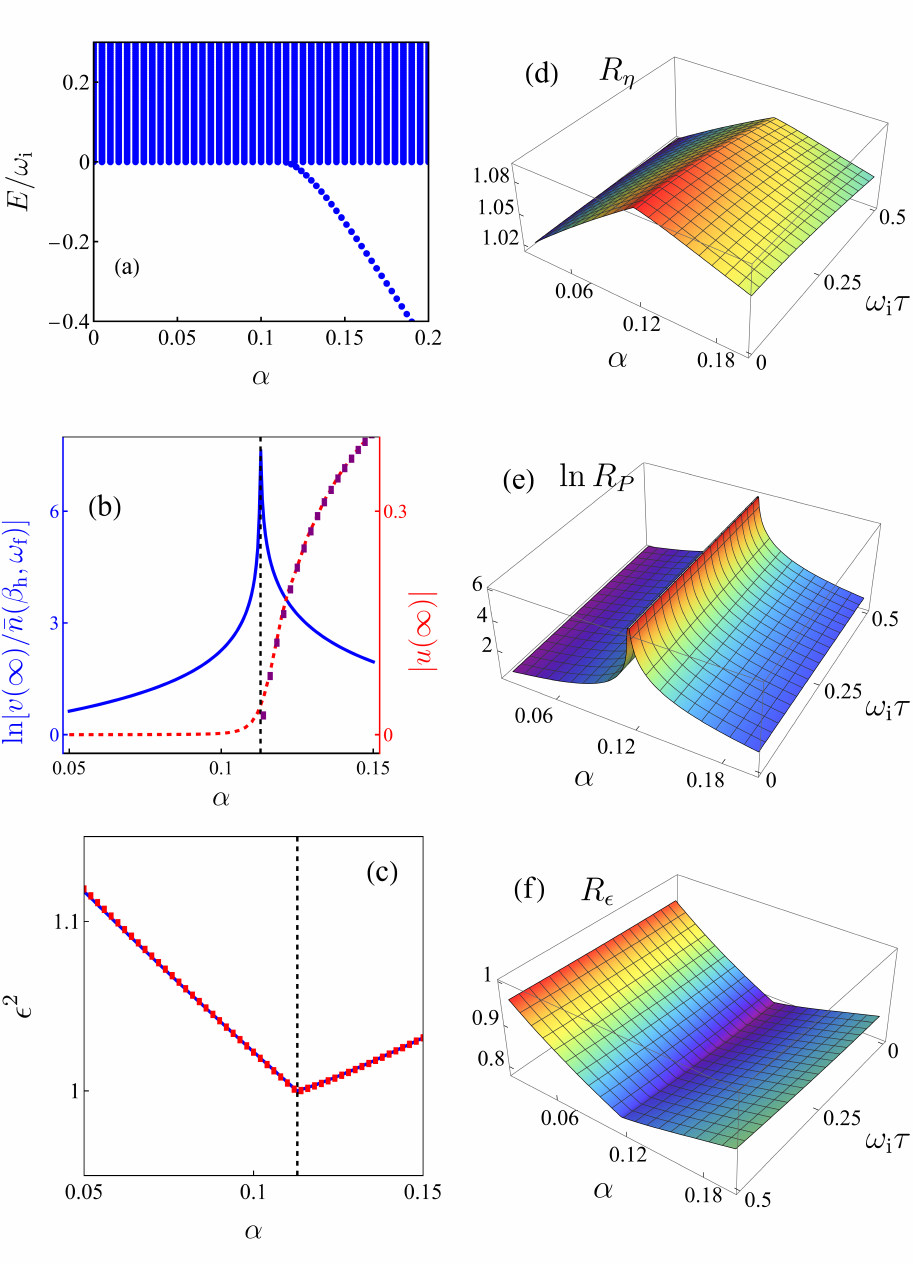}
\caption{(a) Energy spectrum of the total system consisting of the working substance and the hot bath. (b) $v(\infty)/\bar{n}(\beta_{\text{h}},\omega_{\text{f}})$ (blue line) and $|u(\infty)|$ (red dashed line) are plotted as the functions of $\alpha$. The purple rectangles are analytical results $|u(\infty)|=Z$. (c) $\epsilon^{2}$ versus $\alpha$. The red rectangles are numerical results with $\omega_{\text{i}}\tau=100$, while the blue solid line is the adiabatic result of $\epsilon_{\text{adi}}^{2}=1+ 1/v(\infty)$. (d) $R_{\eta}\equiv\langle\eta\rangle/\langle\eta_{\text{BM}}\rangle$ (e) $R_{P}\equiv\langle\mathcal{P}\rangle/\langle\mathcal{P}_{\text{BM}}\rangle$ and (f) $R_{\epsilon}\equiv\epsilon/\epsilon_{\text{M}}$ as the functions of the cycle time $\tau$ and the coupling constant $\alpha$. We use $\omega_{\text{f}}=2\omega_{\text{i}}$, $\beta_{\text{c}}=5\omega_{\text{i}}^{-1}$, $\beta_{\text{h}}=0.1\omega_{\text{i}}^{-1}$, $s=0.5$, and $\omega_{\text{c}}=10\omega_{\text{i}}$.}\label{fig:fig2}
\end{figure}

\section{Non-Markovian case}\label{sec:sec5}
In the general non-Markovian dynamics, one can apply a Laplace transform to linearize Eq.~(\ref{eq:eq3}) into $\tilde{u}(z)=[z+i\omega_{\text{f}}+\int_0^\infty{J(\omega)\over z+i\omega}d\omega]^{-1}$. $u(t)$ is obtained by the inverse Laplace transform of $\tilde{u}(z)$, which can be done by finding its pole from
\begin{equation}\label{eq:eq9}
\bar{y}(\mathcal{E})\equiv\omega_{\text{f}}-\int_0^\infty{\mathcal{J}(\omega)\over\omega-\mathcal{E}}d\omega =\mathcal{E},~(\mathcal{E}=ip).
\end{equation}
It is noted that the roots $\mathcal{E}$ of Eq.~(\ref{eq:eq9}) are just the eigenenergies of $\hat{H}$ in the single-excitation space. Specifically, by expanding the eigenstate as $|\Psi\rangle=(\bar{\alpha}\hat{a}^{\dagger}+\sum_{k}\bar{\beta}_{k}\hat{b}_{k}^{\dagger})|0,\{0_{k}\}\rangle$ and substituting it into $\hat{H}|\Psi\rangle=\mathcal{E}|\Psi\rangle$, we have $(\mathcal{E}-\omega_{\text{f}})\bar{\alpha}=\sum_{k}g_{k}\bar{\beta}_{k}$ and $\bar{\beta}_{k}=g_{k}\bar{\alpha}/(\mathcal{E}-\omega_{k})$. They readily lead to Eq.~(\ref{eq:eq9}). It reveals that, although the Hilbert subspaces with any excitation numbers are involved, the dynamics of the working substance is uniquely determined by the energy-spectrum characteristic of the total system in the single-excitation space. Since $\bar{y}(\mathcal{E})$ is a decreasing function in the regime $\mathcal{E} < 0$, Eq.~(\ref{eq:eq9}) has one isolated root $\mathcal{E}_b$ in this regime provided $\bar{y}(0) < 0$. While $\bar{y}(\mathcal{E})$ is divergent when $\mathcal{E}>0$, Eq.~(\ref{eq:eq9}) has infinite roots in this regime forming a continuous energy band. We call the eigenstate of $\mathcal{E}_b$ bound state. Since an extra band gap is induced by the formation of the bound state, a quantum phase transition occurs in the total system~\cite{Rançon_2013,PhysRevB.84.174301,PhysRevApplied.17.034073,PhysRevA.109.042202}, which has a profound impact on the performance of the Otto engine. This phase transition is reminiscent of the phase transition that occurs in paradigmatic models such as the super-radiance transition in the Dicke model~\cite{PhysRev.93.99,PhysRevE.67.066203,PhysRevLett.107.140402} and the delocalized-localized phase transition in the spin-boson model~\cite{PhysRevLett.99.126801,PhysRevB.95.214308,PhysRevA.105.012431}. 

After the inverse Laplace transform, we obtain~\cite{PhysRevA.103.L010601}
\begin{equation}\label{eq:eq10}
u(t)=Ze^{-i\mathcal{E}_b t}+\int_{0}^{\infty}d\mathcal{E}\Theta(\mathcal{E})\mathcal{J}(\mathcal{E})e^{-i\mathcal{E} t},
\end{equation}
where the first term with $Z=[1+\int_0^\infty{\mathcal{J}(\omega)d\omega\over(\mathcal{E}_b-\omega)^2}]^{-1}$ is contributed by the formed bound state and the second term with $\Theta(\omega)=\{[\omega-\omega_{\text{f}}-\Delta(\omega)]^{2}+[\pi \mathcal{J}(\omega)]^2\}^{-1}$ is from the band energies. Oscillating with time in continuously changing frequencies, the integral tends to zero in the long-time condition due to out-of-phase interference. Thus, if the bound state is absent, then $u(\infty)= 0$ characterizes a complete decoherence, while if the bound state is formed, then $u(\infty)=Ze^{-i\mathcal{E}_b t}$ implies a suppression of decoherence. We can evaluate that the bound state for the Ohmic-family spectral density
is formed if $\omega_{\text{f}}\leq \alpha\omega_c \underline{\Gamma}(s)$, where $\underline{\Gamma}(s)$ is the Euler's gamma function. The long-time form of $v(t)$ is given by~\cite{PhysRevLett.133.050401}
\begin{equation}\label{eq:eq11}
v(\infty)=\int_{0}^{\infty} d\omega \frac{\mathcal{J}(\omega)}{e^{\beta_\text{h}\omega}-1}\Big{[}\Theta(\omega)+{Z^2\over (\omega-\mathcal{E}_b)^2}\Big{]},
\end{equation}
which is sensitive to the details of $\mathcal{J}(\omega)$. Using the quantum reservoir engineering technique~\cite{ER1,Kienzler53,PhysRevA.78.010101}, one can manipulate the value of $v(t)$. The formation of bound state significantly modifies the heat exchange between the the working substance and the bath. This results in the noncanonical distribution of Eq.~(\ref{eq:eqpnm}), which deviates from the standard canonical thermalization \cite{PhysRevE.90.022122,PhysRevA.89.012128,PhysRevResearch.4.023141}. Thus, $\langle\eta\rangle$, $\langle\mathcal{P}\rangle$, and $\epsilon$ are greatly altered. Their exact non-Markovian expressions are provided in the Supplemental Material~\cite{SupplementalMaterial}. In the adiabatic limit $\lim_{\tau_{\ell}\rightarrow\infty}\mathcal{Q}_{\ell}\rightarrow 1$, they reduce to $\langle\eta_{\text{adi}}\rangle=1-\omega_{\text{i}}/\omega_{\text{f}}$, $\langle W_{\text{adi}}\rangle-\langle W^{\text{BM}}_{\text{adi}}\rangle\propto v(\infty)-\bar{n}(\beta_{\text{h}},\omega_{\text{f}})$, and $\epsilon^{2}_{\text{BM,adi}}-\epsilon_{\text{adi}}^{2}\propto 1/\bar{n}(\beta_{\text{h}},\omega_{\text{f}})-1/v(\infty)$~\cite{SupplementalMaterial}. Thus, as long as $v(\infty)>\bar{n}(\beta_{\text{h}},\omega_{\text{f}})$ is satisfied, the non-Markovianity induced noncanonical effect can reduce the fluctuation while increasing the output work. The energy spectrum in Fig.~\ref{fig:fig2}(a) confirms our expectation that the bound state is present when $\alpha\geq\omega_{\text{f}}/[\omega_c \underline{\Gamma}(s)]$. With the formation of the bound state, $|u(t)|$ tends to a finite value $Z$. At the critical point, $v(\infty)$ shows a sharp increment over $\bar{n}(\beta_\text{h},\omega_\text{f})$ (see Fig.~\ref{fig:fig2}(b)). This implies that the synergistic enhancement of the output work and the stability of the engine is most significant near the critical regime. A similar result was experimentally reported~\cite{wei2026experimentaldemonstrationmeasurementfeedbackquantum}. In Fig.~\ref{fig:fig2}(c), via numerical calculations, we verify the adiabatic expression $\epsilon_{\text{adi}}^{2}=1+1/v(\infty)$ and demonstrate that the fluctuation can be reduced near the critical regime.

\begin{figure}
\centering
\includegraphics[angle=0,width=0.475\textwidth]{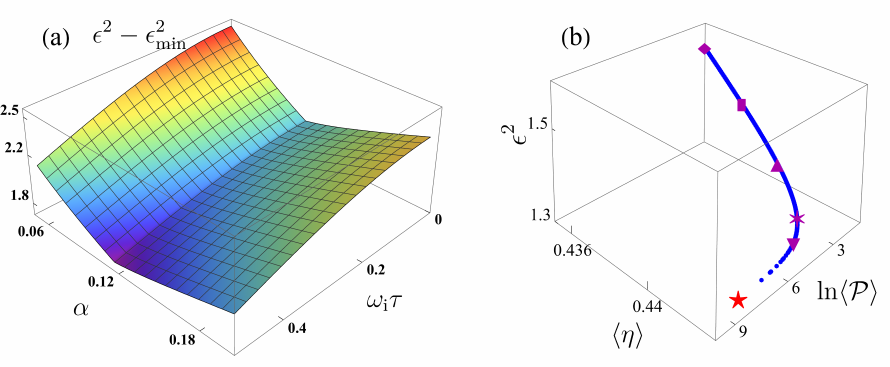}
\caption{(a) $\epsilon^{2}-\epsilon_{\text{text}}^{2}$ is plotted as the functions of $\tau$ and $\alpha$. (b) Trade-off relation among the power $\langle \mathcal{P}\rangle$, the efficiency $\langle \eta\rangle$, and the fluctuation $\epsilon$ with $\omega_{\text{i}}\tau=1$. Different blue points correspond to different coupling strengths. Six representative data are displayed separately: $\alpha=0.05$ (diamond), $\alpha=0.07$ (rectangle), $\alpha=0.09$ (up-triangle), $\alpha=0.12$ (down-triangle), $\alpha=0.13$ (six-pointed star) and $\alpha=0.1128$ (red five-pointed star), which is the critical coupling strength at which the bound state forms. Other parameters are the same as those in Fig.~\ref{fig:fig2}.}\label{fig:fig3}
\end{figure}

The performance can be also boosted in the non-adiabatic case. We quantify the enhancement of the noncanonical effect over the Born-Markov approximate results by introducing $R_{\eta}\equiv\langle\eta\rangle/\langle\eta_{\text{BM}}\rangle$, $R_{P}\equiv\langle\mathcal{P}\rangle/\langle\mathcal{P}_{\text{BM}}\rangle$, and $R_{\epsilon}\equiv\epsilon/\epsilon_{\text{BM}}$. As demonstrated in Fig.~\ref{fig:fig2}(d-f), we find that $R_{\eta}$ and $R_{P}$ are larger than $1$ and their maxima occur at the critical point in which the bound state is formed. Meanwhile, $R_{\epsilon}$ exhibits a minimum value at the critical point. This result means that the efficiency, the power, and the stability of the heat engine are boosted simultaneously. To verify the correctness of Eq.~(\ref{eq:eq12}), we plot $\epsilon^{2}-\epsilon_{\text{min}}^{2}$ in Fig.~\ref{fig:fig3}(a). It exhibits a dip at the critical point, which implies that one can improve the stability at the minimum cost of entropy production with the help of the bound-state effect. As illustrated in Fig.~\ref{fig:fig3}(b), within the bounds permitted by the trade-off relation of Eq.~(\ref{eq:eq12}), the power and efficiency reach their maxima, while the fluctuation reaches its minimum at the critical point, demonstrating that the quantum criticality can simultaneously optimize all three quantities. 

\section{Conclusion and Discussion}\label{sec:sec6}
It should be noted that, although only the harmonic oscillator system is studied as the working substance, the bound-state-enhanced mechanism is hopefully applicable to the Caldeira-Leggett model~\cite{CALDEIRA1983374,CALDEIRA1983587} and the dissipative two-level system. Our proposal is realizable in state-of-the-art quantum-optics experiments because the bound state and its dynamical effect have been observed in circuit quantum electrodynamics platform~\cite{Liu2017} and matter-wave systems~\cite{Kwon2022}. This experimental progress provides strong support to verify our findings. Furthermore, the quantum phase transition in our scheme is induced by the interaction between the working substance and the bath. No criticality occurs once they are decoupled. This is different from almost all the previous quantum critical engines~\cite{Campisi2016,PhysRevE.96.022143,Piccitto_2022,Mukherjee_2021}, in which the working substance itself independently possess the quantum or thermal criticality regardless of whether it interacts with the bath or not. Finally, our work is different from many previous studies of non-Markovian heat engines~\cite{PhysRevE.76.031105,10.1063/5.0192075,Wiedmann_2020,PhysRevA.99.052106,PhysRevA.109.022207,PhysRevE.95.032139,PhysRevResearch.3.023078}. We concentrate on the non-Markovianity induced noncanonical effect on the quantum heat engine. This effect is a novel statical distribution in the long-time condition. However, in previous studies~\cite{PhysRevE.76.031105,10.1063/5.0192075,Wiedmann_2020,PhysRevA.99.052106,HamedaniRaja_2021,PhysRevA.109.022207,PhysRevE.95.032139}, the non-Markovian effect on the averaged efficiency and power of heat engines are discussed from a dynamical perspective in real-time regimes.

In summary, we investigate the influence of non-Markovianity induced noncanonical thermalization on the statistical characteristics of a finite-time quantum Otto cycle. In sharp contrast to many previous studies, it is revealed that the joint distribution function for the efficiency and the power sensitively relies on the structure of the energy spectrum for the composite system consisting of the working substance and the hot bath. By changing the parameters in the spectral density, the average values and the variance of the efficiency and the power are controllable. Moreover, near the quantum critical point, in which a bound state is formed in the single-excitation energy spectrum, we find the performance as well as the stability of the heat engine can be simultaneously boosted at the minimal cost of entropy production. Expanding our understanding of the nonequilibrium quantum thermodynamics, our results lay the foundation for designing an outstanding quantum heat engine.

\section{Acknowledgments}
This work is supported by the National Natural Science Foundation of China (Grants No. 12375015, No. 12275109, No. 92576202, and No. 12247101), the Quantum Science and Technology-National Science and Technology Major Project (Grant No. 2023ZD0300904), the Natural Science Foundation of Gansu Province (Grants No. 26RCKA011 and No. 25JRRA799), and the Fundamental Research Funds for the Central Universities (Grant No. lzujbky-2025-jdzx07).

\bibliography{reference}

\clearpage
\onecolumngrid
\begin{center}

{\large \bf Supplemental Materials for ``Non-Canonical Quantum Otto Engine'' }\\

\vspace{0.3cm}

\end{center}

\begin{center}

{\large \bf Derivation of the generating function}\\

\vspace{0.3cm}

\end{center}

In our study, the working substance is a quantum harmonic oscillator with a time-dependent angular frequency. The corresponding Hamiltonian reads $\hat{H}_{\text{s}}=\frac{1}{2}[\hat{p}^{2}+\omega(t)^{2}\hat{x}^{2}]$. Due to the quadratic characteristic of this Hamiltonian, the Gaussianity of the wave function can be fully preserved during the time evolution. As demonstrated in Refs.~\cite{PhysRevE.77.021128,DEFFNER2010200}, for a given initial Gaussian wave function in the position picture, i.e., $\psi(x_{0},t_{0})=\langle x_{0}|\psi(t_{0})\rangle$ with $\hat{x}|x_{0}\rangle=x_{0}|x_{0}\rangle$, the wave function at $t$ can be expressed as
$\psi(x,t)=\int dx_{0}U(x,t|x_{0},t_{0})\psi(x_{0},t_{0})$, where the propagator $U(x,t|x_{0},t_0)$ is given by~\cite{10.1143/ptp/9.4.381}
\begin{equation}
U(x,t|x_{0},t_0)=\sqrt{\frac{1}{2i\pi X(t)}}\exp\big{\{}\frac{i}{2X(t)}\big{[}\dot{X}(t)x^{2}-2xx_{0}+Y(t)x_{0}^{2}\big{]}\big{\}}
\end{equation}
with $X(t)$ and $Y(t)$ being the solutions of $\ddot{c}(t)+\omega^{2}(t)c(t)=0$ under the boundary conditions $X(t_0)=\dot{Y}(t_0)=0$ and $\dot{X}(t_0)=Y(t_0)=1$. For the linear driving scheme considered in the main text, i.e., $\omega^{2}(t)=\omega^{2}(t_0)-[\omega^{2}(t_{0})-\omega^{2}(\tau_{\ell})]\frac{t}{\tau_{\ell}},~t_{0}\leq t\leq \tau_{\text{c}}$, $X(t)$ and $Y(t)$ have analytical expressions as~\cite{PhysRevE.77.021128}
\begin{eqnarray}
X(t)&=&\frac{\pi\tau_{\ell}^{\frac{1}{3}}}{\{[\omega(t_{0})-\omega(\tau_{\ell})][\omega(t_{0})+\omega(\tau_{\ell})]\}^{\frac{1}{3}}}\big{[}\text{Ai}\big{(}\frac{-\omega^{2}(t_0)\tau_{\ell}^{\frac{2}{3}}}{\{[\omega(t_{0})-\omega(\tau_{\ell})][\omega(t_{0})+\omega(\tau_{\ell})]\}^{\frac{2}{3}}}\big{)}\text{Bi}\big{(}\frac{(t-\tau_{\ell})\omega^{2}(t_0)-t\omega^{2}(\tau_{\ell})}{\tau_{\ell}^{\frac{1}{3}}\{[\omega(t_{0})-\omega(\tau_{\ell})][\omega(t_{0})+\omega(\tau_{\ell})]\}^{\frac{2}{3}}}\big{)}\nonumber\\
&&-\text{Ai}\big{(}\frac{(t-\tau_{\ell})\omega^{2}(t_0)-t\omega^{2}(\tau_{\ell})}{\tau_{\ell}^{\frac{1}{3}}\{[\omega(t_{0})-\omega(\tau_{\ell})][\omega(t_{0})+\omega(\tau_{\ell})]\}^{\frac{2}{3}}}\big{)}\text{Bi}\big{(}\frac{-\omega^{2}(t_0)\tau_{\ell}^{\frac{2}{3}}}{\{[\omega(t_{0})-\omega(\tau_{\ell})][\omega(t_{0})+\omega(\tau_{\ell})]\}^{\frac{2}{3}}}\big{)}\big{]},\\
Y(t)&=&\pi\big{[}\text{Ai}\big{(}\frac{(t-\tau_{\ell})\omega^{2}(t_0)-t\omega^{2}(\tau_{\ell})}{\tau_{\ell}^{\frac{1}{3}}\{[\omega(t_{0})-\omega(\tau_{\ell})][\omega(t_{0})+\omega(\tau_{\ell})]\}^{\frac{2}{3}}}\big{)}\text{Bi}'\big{(}\frac{-\omega^{2}(t_0)\tau_{\ell}^{\frac{2}{3}}}{\{[\omega(t_{0})-\omega(\tau_{\ell})][\omega(t_{0})+\omega(\tau_{\ell})]\}^{\frac{2}{3}}}\big{)}\big{]}\nonumber\\
&&-\text{Ai}'\big{(}\frac{-\omega^{2}(t_0)\tau_{\ell}^{\frac{2}{3}}}{\{[\omega(t_{0})-\omega(\tau_{\ell})][\omega(t_{0})+\omega(\tau_{\ell})]\}^{\frac{2}{3}}}\big{)}\text{Bi}\big{(}\frac{(t-\tau_{\ell})\omega^{2}(t_0)-t\omega^{2}(\tau_{\ell})}{\tau_{\ell}^{\frac{1}{3}}\{[\omega(t_{0})-\omega(\tau_{\ell})][\omega(t_{0})+\omega(\tau_{\ell})]\}^{\frac{2}{3}}}\big{)},
\end{eqnarray}
where $\text{Ai}(z)$ and $\text{Bi}(z)$ are Airy-A and Airy-B functions, respectively, $\text{Ai}'(z)$ and $\text{Bi}'(z)$ denote the first derivatives with respect to $z$. For the compression process, $\omega(t_{0})=\omega_{\text{i}}$ with $t_{0}=0$ and $\omega(\tau_{\text{c}})=\omega_{\text{f}}$. For the expansion process, $\omega(t_{0})=\omega_{\text{f}}$ with $t_{0}=\tau_\text{c}+t_\text{h}$ and $\omega(\tau_{\text{e}})=\omega_{\text{i}}$. Denoting the instantaneous eigenfunctions of the harmonic oscillator as $\phi_{n}(x,t)\equiv\langle x|n,\omega(t)\rangle$, with $\hat{H}_{\text{s}}|n,\omega(t)\rangle=\omega(t)(n+\frac{1}{2})|n,\omega(t)\rangle$, we find that the transition probability $p_{kn}(\tau_{\text{c}})\equiv\big{|}\int \int dx_{0}dx\phi_{n}^{*}(x,\tau_{\text{c}})U(x,\tau_{\text{c}}|x_{0},0)\phi_{k}(x_{0},0)\big{|}^{2}$ can be expressed in terms of the generating function $\mathcal{G}_{\text{c}}(a,b)\equiv\sum_{k,n}a^{k}b^{n}p_{kn}(\tau_{\text{c}})$ as
\begin{equation}
\mathcal{G}_{\text{c}}(a,b)
=\int \int \int \int dx_{0}dy_{0}dxdyU(x,\tau_{\text{c}}|x_{0},0)U^{*}(y,\tau_{\text{c}}|y_{0},0)\sum_{k}a^{k}\phi_{k}(x_{0},0)\phi_{k}^{*}(y_{0},0)\sum_{n}b^{n}\phi_{n}^{*}(x,\tau_{\text{c}})\phi_{n}(y,\tau_{\text{c}}).
\end{equation}
Using the relation~\cite{DEFFNER2010200,10.1143/ptp/9.4.381}
\begin{equation}
\sum_{n=0}^{\infty}a^{n}\phi_{n}(x,t)\phi_{n}^{*}(y,t)=\sqrt{\frac{\omega(t)}{\pi(1-a^{2})}}\exp\big{\{}-\frac{\omega(t)[(1+a^{2})(x^{2}+y^{2})-4a xy]}{2(1-a^{2})}\big{\}},
\end{equation}
and calculating the four-fold Gaussian integral, one finally has a compact expression of $\mathcal{G}_{\text{c}}(a,b)$ as
\begin{equation}
\mathcal{G}_{\text{c}}(a,b)=\sqrt{\frac{2}{\mathcal{Q}_{\text{c}}(1-a^{2})(1-b^{2})+(1+a^{2})(1+b^{2})-4ab}},
\end{equation}
where $\mathcal{Q}_{\text{c}}=\frac{1}{2\omega(t_0)\omega(\tau_{\text{c}})}\{\omega^{2}(t_0)[\omega^{2}(\tau_{\text{c}})X^{2}(\tau_{\text{c}})+\dot{X}^{2}(\tau_{\text{c}})]+[\omega^{2}(\tau_{\text{c}})Y^{2}(\tau_{\text{c}})+\dot{Y}^{2}(\tau_{\text{c}})]\}$ is the non-adiabaticity factor.

With the expression of $\mathcal{G}_{\text{c}}(a,b)$ at hand, one easily finds
\begin{equation} 
\sum_{k}a^{k}\sum_{n}p_{kn}(\tau_{\text{c}})n=\frac{\partial}{\partial b}\mathcal{G}_{\text{c}}(a,b)\big{|}_{b=1}=\frac{\mathcal{Q}_{\text{c}}(1+a)-(1-a)}{2(1-a)^{2}}.
\end{equation}
Expanding the right-hand side of the above equation in powers of $a$ as $\frac{\mathcal{Q}_{\text{c}}(1+a)-(1-a)}{2(1-a)^{2}}=\sum_{k}[\mathcal{Q}_{\text{c}}(k+\frac{1}{2})-\frac{1}{2}]a^{k}$, one immediately gets
\begin{equation}\label{eq:eqr3}
\sum_{n}p_{kn}(\tau_{\text{c}})n=\mathcal{Q}_{\text{c}}\Big{(}k+\frac{1}{2}\Big{)}-\frac{1}{2}.
\end{equation}
In the same manner, we can proceed to obtain
\begin{equation}\label{eq:eqr4}
\sum_{n}p_{kn}(\tau_{\text{c}})n^{2}=\mathcal{X}_{\text{c}}k^{2}+\mathcal{Y}_{\text{c}}k+\mathcal{Z}_{\text{c}},
\end{equation}
where $\mathcal{X}_{\text{c}}=\frac{3}{2}\mathcal{Q}_{\text{c}}^{2}-\frac{1}{2}$, $\mathcal{Y}_{\text{c}}=\frac{3}{2}\mathcal{Q}_{\text{c}}^{2}-\mathcal{Q}_{\text{c}}-\frac{1}{2}$, $\mathcal{Z}_{\text{c}}=\frac{3}{4}\mathcal{Q}_{\text{c}}^{2}-\frac{1}{2}\mathcal{Q}_{\text{c}}-\frac{1}{4}$. One can find the similar equations for the finite-time expansion stroke. Later, these results will be used in the derivations of $\langle W\rangle$, $\langle Q\rangle$ and $\langle W^{2}\rangle$.

\begin{center}

{\large \bf Born-Markov approximate results}\\

\vspace{0.3cm}

\end{center}

In the framework of the Born-Markov approximation, the working substance experiences a canonical thermalization when it absorbs the heat from the hot bath. This results in $p_{nm}^{\text{BM}}(\beta_{\text{h}},\omega_{\text{f}})=\frac{\bar{n}(\beta_{\text{h}},\omega_{\text{f}})^{m}}{[\bar{n}(\beta_{\text{h}},\omega_{\text{f}})+1]^{m+1}}$. Then, the characteristic function $\chi_{\text{BM}}(\xi,\zeta)$ and the corresponding $\langle W_{\text{BM}}\rangle$, $\langle Q_{\text{BM}}\rangle$ and $\langle W_{\text{BM}}^{2}\rangle$ can be analytically derived. In this section, we provide these details. These Born-Markov approximate results will be employed as a benchmark to verify the superiority of our proposed noncanonical quantum Otto heat engine.

Under the Born-Markov approximation, we have
\begin{equation}
P_{\text{BM}}(W,Q)=\frac{2}{\text{csch}(\frac{1}{2}\beta_{\text{c}}\omega_{\text{i}})}\frac{2}{\text{csch}(\frac{1}{2}\beta_{\text{h}}\omega_{\text{f}})}\sum_{k,n,m,l}e^{-\beta_{\text{c}}E_{k}}p_{kn}(\tau_{\text{c}})e^{-\beta_{\text{h}}E_{m}}p_{ml}(\tau_{\text{e}})\delta(W-w_{\text{e}}-w_{\text{c}})\delta(Q-q).
\end{equation}
The corresponding characteristic function can be written as $\chi_{\text{BM}}(\xi,\zeta)=\chi_{\text{c}}^{\text{BM}}(\xi,\zeta)\chi_{\text{e}}^{\text{BM}}(\xi,\zeta)$, where
\begin{equation}
\begin{split}
\chi_{\text{c}}^{\text{BM}}(\xi,\zeta)=&\frac{2\sqrt{a_\text{c}b_\text{c}}}{\text{csch}(\frac{1}{2}\beta_{\text{c}}\omega_{\text{i}})}\sum_{k,n}p_{kn}(\tau_{\text{c}})a_{\text{c}}^{k}b_{\text{c}}^{n}=\frac{2}{\text{csch}(\frac{1}{2}\beta_{\text{c}}\omega_{\text{i}})}\sqrt{\frac{2a_\text{c}b_\text{c}}{\mathcal{Q}_{\text{c}}(1-a_{\text{c}}^{2})(1-b_{\text{c}}^{2})+(1+a_{\text{c}}^{2})(1+b_{\text{c}}^{2})-4a_{\text{c}}b_{\text{c}}}},
\end{split}
\end{equation}
with $a_{\text{c}}\equiv e^{-\omega_{\text{i}}(\beta_{\text{c}}-\xi)}$, $b_{\text{c}}\equiv e^{-\omega_{\text{f}}(\xi-\zeta)}$, and
\begin{equation}
\begin{split}
\chi_{\text{e}}^{\text{BM}}(\xi,\zeta)=&\frac{2\sqrt{a_\text{e}b_\text{e}}}{\text{csch}(\frac{1}{2}\beta_{\text{h}}\omega_{\text{f}})}\sum_{m,l}p_{ml}(\tau_{\text{e}})a_{\text{e}}^{m}b_{\text{e}}^{l}=\frac{2}{\text{csch}(\frac{1}{2}\beta_{\text{h}}\omega_{\text{f}})}\sqrt{\frac{2a_\text{e}b_\text{e}}{\mathcal{Q}_{\text{e}}(1-a_{\text{e}}^{2})(1-b_{\text{e}}^{2})+(1+a_{\text{e}}^{2})(1+b_{\text{e}}^{2})-4a_{\text{e}}b_{\text{e}}}},
\end{split}
\end{equation}
with $a_{\text{e}}\equiv e^{-\omega_{\text{f}}(\beta_{\text{h}}-\xi+\zeta)}$ and $b_{\text{e}}\equiv e^{-\omega_{\text{i}}\xi}$. Here, we have used the definitions of the generating function $\mathcal{G}_{\ell}(a,b)$.

With the analytical expression of $\chi_{\text{BM}}(\xi,\zeta)$ at hand, one immediately has
\begin{eqnarray}
\langle W_{\text{BM}}\rangle&=&-\frac{\partial\ln\chi_{\text{BM}}(\xi,\zeta)}{\partial\xi}\big{|}_{\xi=\zeta=0}=\frac{1}{2}(\mathcal{Q}_{\text{c}}\omega_{\text{f}}-\omega_{\text{i}})\coth\big{(}\frac{\beta_{\text{c}}\omega_{\text{i}}}{2}\big{)}+\frac{1}{2}(\mathcal{Q}_{\text{e}}\omega_{\text{i}}-\omega_{\text{f}})\coth\big{(}\frac{\beta_{\text{h}}\omega_{\text{f}}}{2}\big{)},\label{eq:eqwbm}\\
\langle Q_{\text{BM}}\rangle&=&-\frac{\partial\ln\chi_{\text{BM}}(\xi,\zeta)}{\partial\zeta}\big{|}_{\xi=\zeta=0}=\frac{1}{2}\omega_{\text{f}}\big{[}\coth\big{(}\frac{\beta_{\text{h}}\omega_{\text{f}}}{2}\big{)}-\mathcal{Q}_{\text{c}}\coth\big{(}\frac{\beta_{\text{c}}\omega_{\text{i}}}{2}\big{)}\big{]},\label{eq:eqqbm}\\
\delta^{2}W_{\text{BM}}&=&\frac{\partial^{2}\ln\chi_{\text{BM}}(\xi,\zeta)}{\partial\xi^{2}}\big{|}_{\xi=\zeta=0}=\frac{2e^{\beta_{\text{h}}\omega_{\text{f}}+\beta_{\text{c}}\omega_{\text{i}}}}{(1-e^{\beta_{\text{h}}\omega_{\text{f}}})^{2}(1-e^{\beta_{\text{c}}\omega_{\text{i}}})^{2}} \big{\{}2(\mathcal{Q}_{\text{c}}+\mathcal{Q}_{\text{e}})\omega_{\text{i}}\omega_{\text{f}}\nonumber\\
&&-(\mathcal{Q}^{2}_{\text{c}}+1)\omega^{2}_{\text{f}}-(\mathcal{Q}^{2}_{\text{e}}+1)\omega^{2}_{\text{i}}+\cosh(\beta_{\text{c}}\omega_{\text{i}})\Big{[}\mathcal{Q}_{\text{e}}^{2}\omega_{\text{i}}^{2}-2\mathcal{Q}_{\text{e}}\omega_{\text{i}}\omega_{\text{f}}-(\mathcal{Q}_{\text{c}}^{2}-2)\omega^{2}_{\text{f}}\Big{]}\nonumber\\
&&+\cosh(\beta_{\text{h}}\omega_{\text{f}})\Big{(}\mathcal{Q}_{\text{c}}^{2}\omega_{\text{f}}^{2}-2\mathcal{Q}_{\text{c}}\omega_{\text{i}}\omega_{\text{f}}-(\mathcal{Q}_{\text{e}}^{2}-2)\omega_{\text{i}}^{2}+\cosh(\beta_{\text{c}}\omega_{\text{i}})\Big{[}(\mathcal{Q}_{\text{e}}^{2}-1)\omega_{\text{i}}^{2}+(\mathcal{Q}_{\text{c}}^{2}-1)\omega_{\text{f}}^{2}\Big{]}\Big{)}\big{\}}.\label{eq:eqw2bm}
\end{eqnarray}
It is easily checks that, in the adiabatic limit $\mathcal{Q}_{\ell}=1$, the fluctuation reduces to $\delta^{2}W_{\text{BM}}=\frac{1}{4}(\omega_{\text{f}} - \omega_{\text{i}})^2[\operatorname{csch}^2(\frac{1}{2}\beta_{\text{c}} \omega_{\text{i}}) + \operatorname{csch}^2(\frac{1}{2}\beta_{\text{h}} \omega_{\text{f}})]$, which means $\epsilon^{2}_\text{BM}=[\operatorname{csch}^2(\frac{1}{2}\beta_{\text{c}} \omega_{\text{i}}) + \operatorname{csch}^2(\frac{1}{2}\beta_{\text{h}} \omega_{\text{f}})]/[\coth(\frac{\beta_{\text{c}}\omega_{\text{i}}}{2})-\coth(\frac{\beta_{\text{h}}\omega_{\text{f}}}{2})]^{2}$. All these results are physical reasonbale and reproduce the predictions of previous studies~\cite{PhysRevA.105.022609,Abah_2014,PhysRevE.98.032121}.

\begin{center}

{\large \bf Non-Markovian expressions for $\langle \hat{n}^{1,2}(t)\rangle$}\\

\vspace{0.3cm}

\end{center}

To derive the exact expressions of $\langle W\rangle$, $\langle Q\rangle$ and $\langle W^{2}\rangle$, one needs to evaluate $\langle \hat{n}^{1,2}(t)\rangle\equiv\text{Tr}[\rho_{\text{sb}}(0)e^{i\hat{H}t}\hat{n}^{1,2}e^{i\hat{H}t}]$ with $\hat{n}\equiv \hat{a}^{\dagger}\hat{a}$. In this section, we provide the details of deriving these expressions.

From the total Hamiltonian, one can derive the exact quantum Langevin equation as follows
\begin{equation}
\frac{d}{dt}\hat{a}(t)+i\omega_{\text{f}}\hat{a}(t)+\int_{0}^{t}d\tau\mu(t-\tau)\hat{a}(\tau)=-i\sum_{k}g_{k}\hat{b}_{k}(0)e^{-i\omega_{k}t}.
\end{equation}
The linearity of the above expression allows one to express $\hat{a}(t)=u(t)\hat{a}(0)+\hat{f}(t)$~\cite{PhysRevA.95.033830}, where
\begin{eqnarray}
&&\dot{u}(t)+i\omega_\text{f}u(t)+\int_{0}^{t}d\tau \mu(t-\tau)u(\tau)=0,\\
&&\frac{d}{dt}\hat{f}(t)+i\omega_{\text{f}}\hat{f}(t)+\int_{0}^{t}d\tau \mu(t-\tau)\hat{f}(\tau)=-i\sum_{k}g_{k}\hat{b}_{k}(0)e^{-i\omega_{k}t},\label{ftttd}
\end{eqnarray}
with $u(0)=1$. Equation \eqref{ftttd} can be formally solved as
\begin{equation}\label{eq:eqs19}
\hat{f}(t)=-i\sum_{k}g_{k}\hat{b}_{k}(0)\int_{0}^{t}d\tau e^{-i\omega_{k}\tau}u(t-\tau).
\end{equation}
The expectation value of the number operator $\hat{n}(t)$ is calculated as $\langle \hat{n}(t)\rangle=|u(t)|^{2}\langle\hat{a}^{\dagger}(0)\hat{a}(0)\rangle_{\text{s}}+\langle\hat{f}^{\dagger}(t)\hat{f}(t)\rangle_{\text{b}}$. The substitution of Eq. \eqref{eq:eqs19} results in
\begin{equation}
\begin{split}
\langle\hat{f}^{\dagger}(t)\hat{f}(t)\rangle_{\text{b}}=&\sum_{k}|g_{k}|^{2}\langle\hat{b}^{\dagger}_{k}(0)\hat{b}_{k}(0)\rangle_{\text{b}}\int_{0}^{t}d\tau_{1}\int_{0}^{t}d\tau_{2} u^{*}(t-\tau_{1}) e^{i\omega_{k}(\tau_{1}-\tau_{2})}u(t-\tau_{2})\\
=&\int_{0}^{t}d\tau_{1}\int_{0}^{t}d\tau_{2} u^{*}(\tau_{1}) \nu(\tau_{1}-\tau_{2})u(\tau_{2})=v(t).
\end{split}
\end{equation}
Therefore, we finally arrive at
\begin{equation}
\langle \hat{n}(t)\rangle=|u(t)|^{2}\langle\hat{n}(0)\rangle_{\text{s}}+v(t),
\end{equation}
which reproduces the same result from the exact non-Markovian master equation approach~\cite{PhysRevE.90.022122}. Thus, for a given initial state $\rho_{\text{s}}(0)=\sum_{k}c_{k}|k,\omega_{\text{f}}\rangle\langle k,\omega_{\text{f}}|$, the mean excitation number at $t=\tau$ is given by
\begin{equation}\label{eq:eqa3}
\langle \hat{n}(\tau)\rangle=\sum_{k}c_{k}[|u(\tau)|^{2}k+v(\tau)].
\end{equation}

The mean excitation number is also calculated by $\langle \hat{n}(\tau)\rangle=\text{Tr}[\rho_{\text{s}}(\tau)\hat{n}(0)]$, where $\rho_{\text{s}}(\tau)=\sum_{k}c_{k}\sum_{k'}p_{kk'}(\beta_{\text{h}},\omega_{\text{f}})|k',\omega_{\text{f}}\rangle\langle k',\omega_{\text{f}}|$ with $p_{kk'}(\beta_{\text{h}},\omega_{\text{f}})$ being the transition probability from $k$th eigenstate to the $k'$th eigenstate during the energy exchange process with the hot bath. This means
\begin{equation}\label{eq:eqa4}
\langle \hat{n}(\tau)\rangle=\sum_{k}c_{k}\sum_{k'}p_{kk'}(\beta_{\text{h}},\omega_{\text{f}})k'.
\end{equation}
The equivalence of Eq.~(\ref{eq:eqa3}) and Eq.~(\ref{eq:eqa4}) implies the following relation
\begin{equation}\label{eq:eqr1}
\sum_{k'}p_{kk'}(\beta_{\text{h}},\omega_{\text{f}})k'=|u(\tau)|^{2}k+v(\tau),
\end{equation}
which will be used in our later derivation for the expressions of $\langle W\rangle$ and $\langle Q\rangle$. In the Born-Markov approximation, Eq.~(\ref{eq:eqr1}) reduces to the well-known result for canonical distribution
$\sum_{k'}p_{kk'}(\beta_{\text{h}},\omega_{\text{f}})k'\simeq\sum_{k'}p_{k'}^{\text{BM}}(\beta_{\text{h}},\omega_{\text{f}})k'=\bar{n}(\beta_{\text{h}},\omega_{\text{f}})$.

Next, we derive the expression of $\langle \hat{n}^{2}(t)\rangle$. Using $\hat{a}(t)=u(t)\hat{a}(0)+\hat{f}(t)$, we find
\begin{equation}
\begin{split}
\langle \hat{n}^{2}(t)\rangle=&|u(t)|^{4}\langle\hat{a}^{\dagger}(0)\hat{a}(0)\hat{a}^{\dagger}(0)\hat{a}(0)\rangle_{\text{s}}+|u(t)|^{2}\langle\hat{a}^{\dagger}(0)\hat{a}(0)\rangle_{\text{s}}\langle\hat{f}^{\dagger}(t)\hat{f}(t)\rangle_{\text{b}}\\
&+|u(t)|^{2}\langle\hat{a}^{\dagger}(0)\hat{a}(0)\rangle_{\text{s}}\langle\hat{f}(t)\hat{f}^{\dagger}(t)\rangle_{\text{b}}+|u(t)|^{2}\langle\hat{a}(0)\hat{a}^{\dagger}(0)\rangle_{\text{s}}\langle\hat{f}^{\dagger}(t)\hat{f}(t)\rangle_{\text{b}}\\
&+|u(t)|^{2}\langle\hat{a}^{\dagger}(0)\hat{a}(0)\rangle_{\text{s}}\langle\hat{f}^{\dagger}(t)\hat{f}(t)\rangle_{\text{b}}+\langle\hat{f}^{\dagger}(t)\hat{f}(t)\hat{f}^{\dagger}(t)\hat{f}(t)\rangle_{\text{b}}.
\end{split}
\end{equation}
Here
\begin{equation}\label{eq:eqf1}
\begin{split}
\langle\hat{f}(t)\hat{f}^{\dagger}(t)\rangle_{\text{b}}=&\sum_{k}|g_{k}|^{2}[\bar{n}(\beta_{\text{h}},\omega_{k})+1]\int_{0}^{t}d\tau_{1}\int_{0}^{t}d\tau_{2} u^{*}(\tau_{1}) e^{-i\omega_{k}(\tau_{1}-\tau_{2})}u(\tau_{2})=v(t)+w(t),
\end{split}
\end{equation}
where $w(t)\equiv\int_{0}^{t}d\tau_{1}\int_{0}^{t}d\tau_{2} u^{*}(\tau_{1})\mu(\tau_{1}-\tau_{2}) u(\tau_{2})=1-|u(t)|^2$.
From $\hat{a}(t)=u(t)\hat{a}(0)+\hat{f}(t)$ and Eq.~(\ref{eq:eqf1}), one sees that the commutation relation $[\hat{f}(t),\hat{f}^{\dagger}(t)]=w(t)$. Using this relation, one has
$\hat{f}^{\dagger}(t)\hat{f}(t)\hat{f}^{\dagger}(t)\hat{f}(t)=\hat{f}^{\dagger}(t)\hat{f}^{\dagger}(t)\hat{f}(t)\hat{f}(t)+w(t)\hat{f}^{\dagger}(t)\hat{f}(t)$. From Eq.~(\ref{eq:eqs19}), we find~\cite{PhysRevA.95.033830}
\begin{equation}\label{eq:eqs31}
\begin{split}
\langle\hat{f}^{\dagger}(t)\hat{f}^{\dagger}(t)\hat{f}(t)\hat{f}(t)\rangle_{\text{b}}
=&2\big{\{}\sum_{k}|g_{k}|^{2}\bar{n}(\beta_{\text{b}},\omega_{k})\int_{0}^{t}d\tau_{1}\int_{0}^{t}d\tau_{2} u^{*}(\tau_{1}) e^{-i\omega_{k}(\tau_{1}-\tau_{2})}u(\tau_{2})\big{\}}\\
&\times\big{\{}\sum_{k'}|g_{k'}|^{2}\bar{n}(\beta_{\text{b}},\omega_{k'})\int_{0}^{t}d\tau_{1}\int_{0}^{t}d\tau_{2} u^{*}(\tau_{1}) e^{-i\omega_{k'}(\tau_{1}-\tau_{2})}u(\tau_{2})\big{\}}=2v^{2}(t),
\end{split}
\end{equation}
where we have used $\langle\hat{b}_{k}^{\dagger}(0)\hat{b}_{k}^{\dagger}(0)\hat{b}_{k}(0)\hat{b}_{k}(0)\rangle_{\text{b}}=2\bar{n}(\beta_{\text{h}},\omega_{k})^{2}$. Combining Eq.~(\ref{eq:eqf1}) and Eq.~(\ref{eq:eqs31}), we finally arrive at
\begin{equation}
\begin{split}
\langle \hat{n}^{2}(t)\rangle=&|u(t)|^{4}\langle\hat{n}^{2}(0)\rangle_{\text{s}}+|u(t)|^{2}[4v(t)+w(t)]\langle\hat{n}(0)\rangle_{\text{s}}+2v(t)^{2}+v(t)w(t)+|u(t)|^{2}v(t).
\end{split}
\end{equation}
Using the above result, one can prove that
\begin{equation}\label{eq:eqr2}
\sum_{k'}p_{kk'}(\beta_{\text{h}},\omega_{\text{f}})k'^{2}=\mathcal{A}(\tau)k^{2}+\mathcal{B}(\tau)k+\mathcal{C}(\tau),
\end{equation}
where $\mathcal{A}(\tau)=|u(\tau)|^{4}$, $\mathcal{B}(\tau)=|u(\tau)|^{2}[4v(\tau)+w(\tau)]$, and $\mathcal{C}(\tau)=2v(\tau)^{2}+v(\tau)w(\tau)+|u(\tau)|^{2}v(\tau)$. The above relation will be used in our later derivations for the expression of $\langle W^{2}\rangle$. In the Born-Markov approximation, $w_{\text{BM}}(\infty)=1$, Eq.~(\ref{eq:eqr2}) reduces to the well-known result for the canonical distribution
\begin{equation}
\sum_{k'}p_{kk'}(\beta_{\text{h}},\omega_{\text{f}})k'^{2}\simeq\sum_{k'}p_{k'}^{\text{BM}}(\beta_{\text{h}},\omega_{\text{f}})k'^{2}=2\bar{n}(\beta_{\text{h}},\omega_{\text{f}})^{2}+\bar{n}(\beta_{\text{h}},\omega_{\text{f}}).
\end{equation}

\begin{center}

{\large \bf Expressions of $\langle W\rangle$ and $\langle Q\rangle$}\\

\vspace{0.3cm}

\end{center}

In this section, we derive the exact non-Markovian expressions of $\langle W\rangle$ and $\langle Q\rangle$, from which the mean value of the efficiency and the power can be accordingly obtained. The mean work is
\begin{equation}\label{eq:eqw1}
\begin{split}
\langle W\rangle=&\sum_{k}p_{k}(\beta_{\text{c}},\omega_{\text{i}})\sum_{n}p_{kn}(\tau_{\text{c}})E_{n}+\sum_{k}p_{k}(\beta_{\text{c}},\omega_{\text{i}})\sum_{n}p_{kn}(\tau_{\text{c}})\sum_{m}p_{nm}(\beta_{\text{h}},\omega_{\text{f}})\sum_{l}p_{ml}(\tau_{\text{e}})E_{l}\\
&-\sum_{k}p_{k}(\beta_{\text{c}},\omega_{\text{i}})E_{k}-\sum_{k}p_{k}(\beta_{\text{c}},\omega_{\text{i}})\sum_{n}p_{kn}(\tau_{\text{c}})\sum_{m}p_{nm}(\beta_{\text{h}},\omega_{\text{f}})E_{m}.
\end{split}
\end{equation}
Notice that the reduced density matrix of the working substance at the end of the four strokes are given by
\begin{eqnarray}
\rho_{\text{A}}&=&\sum_{k}p_{k}(\beta_{\text{c}},\omega_{\text{i}})|k,\omega_{\text{i}}\rangle\langle k,\omega_{\text{i}}|,\\
\rho_{\text{B}}&=&\sum_{k}p_{k}(\beta_{\text{c}},\omega_{\text{i}})\sum_{n}p_{kn}(\tau_{\text{c}})|n,\omega_{\text{f}}\rangle\langle n,\omega_{\text{f}}|,\\
\rho_{\text{C}}&=&\sum_{k}p_{k}(\beta_{\text{c}},\omega_{\text{i}})\sum_{n}p_{kn}(\tau_{\text{c}})\sum_{m}p_{nm}(\beta_{\text{h}},\omega_{\text{f}})|m,\omega_{\text{f}}\rangle\langle m,\omega_{\text{f}}|,\\
\rho_{\text{D}}&=&\sum_{k}p_{k}(\beta_{\text{c}},\omega_{\text{i}})\sum_{n}p_{kn}(\tau_{\text{c}})\sum_{m}p_{nm}(\beta_{\text{h}},\omega_{\text{f}})\sum_{l}p_{ml}(\tau_{\text{e}})|l,\omega_{\text{i}}\rangle\langle l,\omega_{\text{i}}|,
\end{eqnarray}
Eq.~(\ref{eq:eqw1}) is recast into
$\langle W\rangle=\langle \hat{H}_{\text{B}}\rangle-\langle \hat{H}_{\text{A}}\rangle+\langle \hat{H}_{\text{D}}\rangle-\langle \hat{H}_{\text{C}}\rangle$, where $\langle \hat{H}_{\text{j}}\rangle=\text{Tr}(\rho_{\text{s}}\hat{H}_{\text{j}})$. The above equation is physically reasonable because $\langle \hat{H}_{\text{B}}\rangle-\langle \hat{H}_{\text{A}}\rangle=\langle W_{\text{c}}\rangle$ can be regarded as the mean work done during the compression process and $\langle \hat{H}_{\text{D}}\rangle-\langle \hat{H}_{\text{C}}\rangle=\langle W_{\text{e}}\rangle$ is the mean work done during the expansion process. Similarly, one sees $\langle Q\rangle=\langle \hat{H}_{\text{C}}\rangle-\langle \hat{H}_{\text{B}}\rangle$, which reflects the average energy exchanged from the hot bath. Next, we provide the details of deriving $\langle \hat{H}_{\text{j}}\rangle$.

At the beginning of the quantum Otto cycle (at the stage A), the state of the working substance is a thermal Gibbs state $\rho_{\text{A}}=\sum_{k}p_{k}(\beta_{\text{c}},\omega_{\text{i}})|k,\omega_{\text{i}}\rangle\langle k,\omega_{\text{i}}|$ where $p_{k}(\beta_{\text{c}},\omega_{\text{i}})=\bar{n}(\beta_{\text{c}},\omega_{\text{i}})^{k}/[\bar{n}(\beta_{\text{c}},\omega_{\text{i}})+1]^{k+1}$ with $\bar{n}(\beta_{\text{c}},\omega_{\text{i}})=1/[\exp(\beta_{\text{c}}\omega_{\text{i}})-1]$ being the mean excitation number. Then the
average energy of the work medium is
\begin{equation}
\begin{split}
\langle\hat{H}_{\text{A}}\rangle=&\omega_{\text{i}}\sum_{k}p_{k}(\beta_{\text{c}},\omega_{\text{i}})\big{(}k+\frac{1}{2}\big{)}=\omega_{\text{i}}\big{[}\bar{n}(\beta_{\text{c}},\omega_{\text{i}})+\frac{1}{2}\big{]}.
\end{split}
\end{equation}
Next, the working substance undergoes a finite-time compression process in which the frequency changes from $\omega_{\text{i}}$ to $\omega_{\text{f}}$. Its state at the stage B is $\rho_{\text{B}}=\sum_{k}p_{k}(\beta_{\text{c}},\omega_{\text{i}})\sum_{n}p_{kn}(\tau_{\text{c}})|n,\omega_{\text{f}}\rangle\langle n,\omega_{\text{f}}|$ where $p_{kn}(\tau_{\text{c}})$ is the transition probability from the $k$-th eigenstate $|k,\omega_{\text{i}}\rangle$ at $t=0$ to the $n$-th eigenstate $|n,\omega_{\text{f}}\rangle$ at $t=\tau_{\text{c}}$. Thus the average energy of $\hat{H}_{\text{s}}$ at stage B is
\begin{equation}
\begin{split}
\langle \hat{H}_{\text{B}}\rangle=&\sum_{k}p_{k}(\beta_{\text{c}},\omega_{\text{i}})\sum_{n}p_{kn}(\tau_{\text{c}})\langle n,\omega_{\text{f}}|\hat{H}_{\text{s}}(\omega_{\text{f}})| n,\omega_{\text{f}}\rangle\\
=&\omega_{\text{f}}\sum_{k}p_{k}(\beta_{\text{c}},\omega_{\text{i}})\big{[}\big{(}k+\frac{1}{2}\big{)}\mathcal{Q}_{\text{c}}-\frac{1}{2}\big{]}+\frac{1}{2}\omega_{\text{f}}=\omega_{\text{f}}\mathcal{Q}_{\text{c}}\big{[}\bar{n}(\beta_{\text{c}},\omega_{\text{i}})+\frac{1}{2}\big{]},
\end{split}
\end{equation}
where Eq.~(\ref{eq:eqr3}) has been used.

In the third stroke, the working substance interacts with the hot bath and undergoes a non-Markovian relaxation process. At the end of relaxation (the stage C), we have $\rho_{\text{C}}=\sum_{k}p_{k}(\beta_{\text{c}},\omega_{\text{i}})\sum_{n}p_{kn}(\tau_{\text{c}})\sum_{m}p_{nm}(\beta_{\text{h}},\omega_{\text{f}})|m,\omega_{\text{f}}\rangle\langle m,\omega_{\text{f}}|$ where $p_{nm}(\beta_{\text{h}},\omega_{\text{f}})$ is the transition probability from the $n$-th eigenstate $|n,\omega_{\text{f}}\rangle$ to the $m$-th eigenstate $|m,\omega_{\text{f}}\rangle$ during the heat exchange process with the fixed frequency $\omega(t)=\omega_{\text{f}}$. The average energy at the end of the stroke is then given by
\begin{equation}
\begin{split}
\langle \hat{H}_{\text{C}}\rangle=&\sum_{k}p_{k}(\beta_{\text{c}},\omega_{\text{i}})\sum_{n}p_{kn}(\tau_{\text{c}})\sum_{m}p_{nm}(\beta_{\text{h}},\omega_{\text{f}})\langle m,\omega_{\text{f}}|\hat{H}_{\text{s}}(\omega_{\text{f}})|m,\omega_{\text{f}}\rangle\\
=&\omega_{\text{f}}\sum_{k}p_{k}(\beta_{\text{c}},\omega_{\text{i}})\sum_{n}p_{kn}(\tau_{\text{c}})\big{\{}[|u(\infty)|^{2}n+v(\infty)]+\frac{1}{2}\big{\}}=\omega_{\text{f}}\big{\{}|u(\infty)|^{2}\big{[}\big{(}\bar{n}(\beta_{\text{c}},\omega_{\text{i}})+\frac{1}{2}\big{)}\mathcal{Q}_{\text{c}}-\frac{1}{2}\big{]}+v(\infty)+\frac{1}{2}\big{\}},
\end{split}
\end{equation}
where Eqs.~(\ref{eq:eqr1}) and (\ref{eq:eqr3}) have been used. During the fourth stroke, the working substance undergoes a finite-time expansion process in which the frequency changes from $\omega_{\text{f}}$ back to $\omega_{\text{i}}$. At the end of the fourth stroke, the reduced density matrix is $\rho_{\text{D}}$ and the corresponding average energy is
\begin{equation}
\begin{split}
\langle \hat{H}_{\text{D}}\rangle=&\sum_{k}p_{k}(\beta_{\text{c}},\omega_{\text{i}})\sum_{n}p_{kn}(\tau_{\text{c}})\sum_{m}p_{nm}(\beta_{\text{h}},\omega_{\text{f}})\sum_{l}p_{ml}(\tau_{\text{e}})\langle l,\omega_{\text{i}}|\hat{H}_{\text{s}}(\omega_{\text{i}})| l,\omega_{\text{i}}\rangle\\
=&\omega_{\text{i}}\mathcal{Q}_{\text{e}}\sum_{k}p_{k}(\beta_{\text{c}},\omega_{\text{i}})\sum_{n}p_{kn}(\tau_{\text{c}})\big{[}|u(\infty)|^{2}n+v(\infty)+\frac{1}{2}\big{]}=\omega_{\text{i}}\mathcal{Q}_{\text{e}}\big{\{}|u(\infty)|^{2}\big{[}\mathcal{Q}_{\text{c}}\big{(}\bar{n}(\beta_{\text{c}},\omega_{\text{i}})+\frac{1}{2}\big{)}-\frac{1}{2}\big{]}+v(\infty)+\frac{1}{2}\big{\}}.
\end{split}
\end{equation}

Under the Born-Markov approximation, as demonstrated in the main text, one has $u_{\text{BM}}(\infty)=0$ and $v_{\text{BM}}(\infty)=\bar{n}(\beta_{\text{h}},\omega_{\text{f}})$, which results in
\begin{equation}
\langle \hat{H}_{\text{C}}^{\text{BM}}\rangle=\omega_{\text{f}}\big{[}\bar{n}(\beta_{\text{h}},\omega_{\text{f}})+\frac{1}{2}\big{]},~~~\langle \hat{H}^{\text{BM}}_{\text{D}}\rangle=\omega_{\text{f}}\mathcal{Q}_{\text{e}}\big{[}\bar{n}(\beta_{\text{h}},\omega_{\text{f}})+\frac{1}{2}\big{]}.
\end{equation}
Then, one easily find that our exact expressions of $\langle W\rangle$ and $\langle Q\rangle$ in conformity with these of the previous Born-Markov approximate results displayed in Eq.~(\ref{eq:eqwbm}) and Eq.~(\ref{eq:eqqbm}).

On the other hand, in the adiabatic limit $\mathcal{Q}_{\ell}=1$, we find
\begin{equation}
\langle W_{\text{adi}}\rangle=(\omega_{\text{f}}-\omega_{\text{i}})[\bar{n}(\beta_{\text{c}},\omega_{\text{i}})(1-|u(\infty)|^{2})-v(\infty)],~~~\langle Q_{\text{adi}}\rangle=-\omega_{\text{f}}[\bar{n}(\beta_{\text{c}},\omega_{\text{i}})(1-|u(\infty)|^{2})-v(\infty)],
\end{equation} 
which fully recovers the Otto efficiency $\langle\eta_{\text{adi}}\rangle=\eta_{\text{Otto}}=1-\omega_{\text{i}}/\omega_{\text{f}}$. The corresponding power reads
\begin{equation}
\langle \mathcal{P}_{\text{adi}}\rangle=\frac{\omega_{\text{f}}-\omega_{\text{i}}}{2\tau}[v(\infty)-\bar{n}(\beta_{\text{c}},\omega_{\text{i}})(1-|u(\infty)|^{2})],
\end{equation} 
which vanishes in the limit of $\tau\rightarrow\infty$. Compared with the Born-Markov approximate result, one sees
\begin{equation}
\langle W_{\text{adi}}\rangle-\langle W^{\text{BM}}_{\text{adi}}\rangle=(\omega_{\text{f}}-\omega_{\text{i}})[v(\infty)+\bar{n}(\beta_{\text{c}},\omega_{\text{i}})|u(\infty)|^{2}-\bar{n}(\beta_{\text{h}},\omega_{\text{f}})].
\end{equation} 
Thus, as long as the condition of
$v(\infty)\gg\bar{n}(\beta_{\text{h}},\omega_{\text{f}})$ is satisfied, the non-Markovianity induced noncanonical effect can boost the output work.

\begin{center}

{\large \bf Expression of $\langle W^{2}\rangle$}\\

\vspace{0.3cm}

\end{center}
Now, we provide the non-Markovian expression of $\langle W^{2}\rangle$. Remembering $W=E_n-E_k+E_l-E_m$, we have
\begin{equation}
\begin{split}
\langle W^{2}\rangle=&\sum_{k}p_{k}(\beta_{\text{c}},\omega_{\text{i}})\sum_{n}p_{kn}(\tau_{\text{c}})\sum_{m}p_{nm}(\beta_{\text{h}},\omega_{\text{f}})\sum_{l}p_{ml}(\tau_{\text{e}})\big{[}E_{n}^{2}+E_{l}^{2}+E_{k}^{2}+E_{m}^{2}\\
&-2\big{(}E_{n}E_{k}+E_{l}E_{m}+E_{n}E_{m}+E_{k}E_{l}-E_{n}E_{l}-E_{k}E_{m}\big{)}\big{]}\\
=&\Pi_{1}+\Pi_{2}+\Pi_{3}+\Pi_{4}-2(\Pi_{5}+\Pi_{6}+\Pi_{7}+\Pi_{8}-\Pi_{9}-\Pi_{10}).
\end{split}
\end{equation}
There are ten terms $\Pi_{1-10}$ need to be evaluated. The first term is
\begin{equation}
\begin{split}
\Pi_{1}&=\omega_{\text{f}}^{2}\sum_{k}p_{k}(\beta_{\text{c}},\omega_{\text{i}})\sum_{n}p_{kn}(\tau_{\text{c}})\big{(}n^{2}+n+\frac{1}{4}\big{)}\\
&=\omega_{\text{f}}^{2}\big{\{}\mathcal{X}_{\text{c}}[2\bar{n}^{2}(\beta_{\text{c}},\omega_{\text{i}})+\bar{n}(\beta_{\text{c}},\omega_{\text{i}})]+(\mathcal{Y}_{\text{c}}+\mathcal{Q}_{\text{c}})\bar{n}(\beta_{\text{c}},\omega_{\text{i}})+\mathcal{Z}_{\text{c}}+\frac{1}{2}\mathcal{Q}_{\text{c}}-\frac{1}{4}\big{\}},
\end{split}
\end{equation}
where Eqs.~(\ref{eq:eqr3}) and (\ref{eq:eqr4}) have been used. The second term is
\begin{equation}
\begin{split}
\Pi_{2}=&\omega_{\text{i}}^{2}\sum_{k}p_{k}(\beta_{\text{c}},\omega_{\text{i}})\sum_{n}p_{kn}(\tau_{\text{c}})\sum_{m}p_{nm}(\beta_{\text{h}},\omega_{\text{f}})\big{[}\mathcal{X}_{\text{e}}m^{2}+(\mathcal{Y}_{\text{e}}+\mathcal{Q}_{\text{e}})m+\mathcal{Z}_{\text{e}}+\frac{1}{2}\mathcal{Q}_{\text{e}}-\frac{1}{4}\big{]}\\
=&\omega_{\text{i}}^{2}\sum_{k}p_{k}(\beta_{\text{c}},\omega_{\text{i}})\sum_{n}p_{kn}(\tau_{\text{c}})\big{\{}\mathcal{X}_{\text{e}}(\mathcal{A} n^{2}+\mathcal{B} n+\mathcal{C})+(\mathcal{Y}_{\text{e}}+\mathcal{Q}_{\text{e}})\Big{[}|u(\infty)|^{2}n+v(\infty)\Big{]}+\mathcal{Z}_{\text{e}}+\frac{1}{2}\mathcal{Q}_{\text{e}}-\frac{1}{4}\big{\}},
\end{split}
\end{equation}
where Eqs.~(\ref{eq:eqr1}) and ~(\ref{eq:eqr2}) have been used and $\mathcal{A}\equiv\mathcal{A}(\infty)$, $\mathcal{B}\equiv\mathcal{B}(\infty)$, and $\mathcal{C}\equiv\mathcal{C}(\infty)$. Using Eq.~(\ref{eq:eqr3}) and Eq.~(\ref{eq:eqr4}) again, it is finally simplified to
\begin{equation}
\begin{split}
\Pi_{2}=&\omega_{\text{i}}^{2}\big{\{}\mathcal{X}_{\text{c}}\mathcal{X}_{\text{e}}\mathcal{A} [2\bar{n}^{2}(\beta_{\text{c}},\omega_{\text{i}})+\bar{n}(\beta_{\text{c}},\omega_{\text{i}})]+\big{(}\mathcal{X}_{\text{e}}\mathcal{Y}_{\text{c}}\mathcal{A}+\mathcal{Q}_{\text{c}}\Big{[}\mathcal{X}_{\text{e}}\mathcal{B}+ (\mathcal{Y}_{\text{e}}+\mathcal{Q}_{\text{e}})|u(\infty)|^{2}\Big{]}\big{)}\bar{n}(\beta_{\text{c}},\omega_{\text{i}})\\
&+\mathcal{X}_{\text{e}}\mathcal{Z}_{\text{c}}\mathcal{A}+\frac{1}{2}\Big{[}\mathcal{X}_{\text{e}}\mathcal{B}+ (\mathcal{Y}_{\text{e}}+\mathcal{Q}_{\text{e}})|u(\infty)|^{2}\Big{]}(\mathcal{Q}_{\text{c}}-1)+\mathcal{X}_{\text{e}}\mathcal{C}+(\mathcal{Y}_{\text{e}}+\mathcal{Q}_{\text{e}})v(\infty)+\mathcal{Z}_{\text{e}}+\frac{1}{2}\mathcal{Q}_{\text{e}}-\frac{1}{4}\big{\}}.
\end{split}
\end{equation}

Using the same method, all the other terms can be exactly derived. Their expressions are given by
\begin{eqnarray}
\Pi_{3}&=&\omega_{\text{i}}^{2}\big{[}2\bar{n}^{2}(\beta_{\text{c}},\omega_{\text{i}})+2\bar{n}(\beta_{\text{c}},\omega_{\text{i}})+\frac{1}{4}\big{]},\\
\Pi_{4}&=&\omega_{\text{f}}^{2}\big{\{}\mathcal{A}\mathcal{X}_{\text{c}}[2\bar{n}^{2}(\beta_{\text{c}},\omega_{\text{i}})+\bar{n}(\beta_{\text{c}},\omega_{\text{i}})]+\Big{[}\mathcal{A}\mathcal{Y}_{\text{c}}+(\mathcal{B}+|u(\infty)|^{2})\mathcal{Q}_{\text{c}}\Big{]}\bar{n}(\beta_{\text{c}},\omega_{\text{i}})\\
&&+\mathcal{A} \mathcal{Z}_{\text{c}}+\frac{1}{2}(\mathcal{B}+|u(\infty)|^{2})(\mathcal{Q}_{\text{c}}-1)+\mathcal{C}+v(\infty)+\frac{1}{4}\big{\}},\\
\Pi_{5}&=&\omega_{\text{i}}\omega_{\text{f}}\mathcal{Q}_{\text{c}}\big{[}2\bar{n}^{2}(\beta_{\text{c}},\omega_{\text{i}})+2\bar{n}(\beta_{\text{c}},\omega_{\text{i}})+\frac{1}{4}\big{]},\\
\Pi_{6}&=&\omega_{\text{i}}\omega_{\text{f}}\mathcal{Q}_{\text{e}}\big{\{}\mathcal{A}\Big{(}\mathcal{X}_{\text{c}}[2\bar{n}^{2}(\beta_{\text{c}},\omega_{\text{i}})+\bar{n}(\beta_{\text{c}},\omega_{\text{i}})]+\mathcal{Y}_{\text{c}}\bar{n}(\beta_{\text{c}},\omega_{\text{i}})+\mathcal{Z}_{\text{c}}\Big{)}\\
&&+[\mathcal{B}+|u(\infty)|^{2}]\big{[}\big{(}\bar{n}(\beta_{\text{c}},\omega_{\text{i}})+\frac{1}{2}\big{)}\mathcal{Q}_{\text{c}}-\frac{1}{2}\big{]}+\mathcal{C}+v(\infty)+\frac{1}{4}\big{\}},\\
\Pi_{7}&=&\omega_{\text{f}}^{2}\big{\{}|u(\infty)|^{2}\Big{(}\mathcal{X}_{\text{c}}[2\bar{n}^{2}(\beta_{\text{c}},\omega_{\text{i}})+\bar{n}(\beta_{\text{c}},\omega_{\text{i}})]+\mathcal{Y}_{\text{c}}\bar{n}(\beta_{\text{c}},\omega_{\text{i}})+\mathcal{Z}_{\text{c}}\Big{)}\\
&&+\big{[}\frac{1}{2}|u(\infty)|^{2}+v(\infty)+\frac{1}{2}\big{]}\big{[}\big{(}\bar{n}(\beta_{\text{c}},\omega_{\text{i}})+\frac{1}{2}\big{)}\mathcal{Q}_{\text{c}}-\frac{1}{2}\big{]}+\frac{1}{2}v(\infty)+\frac{1}{4}\big{\}},\\
\Pi_{8}&=&\omega_{\text{i}}^{2}\mathcal{Q}_{\text{e}}\big{\{}|u(\infty)|^{2}\mathcal{Q}_{\text{c}}[2\bar{n}^{2}(\beta_{\text{c}},\omega_{\text{i}})+\bar{n}(\beta_{\text{c}},\omega_{\text{i}})]+\big{[}\frac{1}{2}|u(\infty)|^{2}(2\mathcal{Q}_{\text{c}}-1)+v(\infty)+\frac{1}{2}\big{]}\bar{n}(\beta_{\text{c}},\omega_{\text{i}})\\
&&+\frac{1}{4}|u(\infty)|^{2}(\mathcal{Q}_{\text{c}}-1)+\frac{1}{2}v(\infty)+\frac{1}{4}\big{\}},\\
\Pi_{9}&=&\omega_{\text{i}}\omega_{\text{f}}\mathcal{Q}_{\text{e}}\big{\{}|u(\infty)|^{2}\Big{(}\mathcal{X}_{\text{c}}[2\bar{n}^{2}(\beta_{\text{c}},\omega_{\text{i}})+\bar{n}(\beta_{\text{c}},\omega_{\text{i}})]+\mathcal{Y}_{\text{c}}\bar{n}(\beta_{\text{c}},\omega_{\text{i}})+\mathcal{Z}_{\text{c}}\Big{)}\\
&&+\big{[}\frac{1}{2}|u(\infty)|^{2}+v(\infty)+\frac{1}{2}\big{]}\big{[}\big{(}\bar{n}(\beta_{\text{c}},\omega_{\text{i}})+\frac{1}{2}\big{)}\mathcal{Q}_{\text{c}}-\frac{1}{2}\big{]}+\frac{1}{2}v(\infty)+\frac{1}{4}\big{\}},\\
\Pi_{10}&=&\omega_{\text{i}}\omega_{\text{f}}\big{\{}|u(\infty)|^{2}\mathcal{Q}_{\text{c}}[2\bar{n}^{2}(\beta_{\text{c}},\omega_{\text{i}})+\bar{n}(\beta_{\text{c}},\omega_{\text{i}})]+\big{[}\frac{1}{2}|u(\infty)|^{2}(2\mathcal{Q}_{\text{c}}-1)+v(\infty)+\frac{1}{2}\big{]}\bar{n}(\beta_{\text{c}},\omega_{\text{i}})\\
&&+\frac{1}{4}|u(\infty)|^{2}(\mathcal{Q}_{\text{c}}-1)+\frac{1}{2}v(\infty)+\frac{1}{4}\big{\}}.
\end{eqnarray}

Under the Born-Markov approximation, one has
\begin{eqnarray}
\Pi_{2}^{\text{BM}}&=&\omega_{\text{i}}^{2}\big{[}2\mathcal{X}_{\text{e}}\bar{n}(\beta_{\text{h}},\omega_{\text{f}})^{2}+(\mathcal{X}_{\text{e}}+\mathcal{Y}_{\text{e}}+\mathcal{Q}_{\text{e}})\bar{n}(\beta_{\text{h}},\omega_{\text{f}})+\mathcal{Z}_{\text{e}}+\frac{1}{2}\mathcal{Q}_{\text{e}}-\frac{1}{4}\big{]},\\
\Pi_{4}^{\text{BM}}&=&\omega_{\text{f}}^{2}\big{[}2\bar{n}(\beta_{\text{h}},\omega_{\text{f}})^{2}+2\bar{n}(\beta_{\text{h}},\omega_{\text{f}})+\frac{1}{4}\big{]},\\
\Pi_{6}^{\text{BM}}&=&\omega_{\text{i}}\omega_{\text{f}}\mathcal{Q}_{\text{e}}\big{[}2\bar{n}(\beta_{\text{h}},\omega_{\text{f}})^{2}+2\bar{n}(\beta_{\text{h}},\omega_{\text{f}})+\frac{1}{4}\big{]},\\
\Pi_{7}^{\text{BM}}&=&\omega_{\text{f}}^{2}\big{\{}\big{[}\bar{n}(\beta_{\text{h}},\omega_{\text{f}})+\frac{1}{2}\big{]}\big{[}\big{(}\bar{n}(\beta_{\text{c}},\omega_{\text{i}})+\frac{1}{2}\big{)}\mathcal{Q}_{\text{c}}-\frac{1}{2}\big{]}+\frac{1}{2}\bar{n}(\beta_{\text{h}},\omega_{\text{f}})+\frac{1}{4}\big{\}},\\
\Pi_{8}^{\text{BM}}&=&\omega_{\text{i}}^{2}\mathcal{Q}_{\text{e}}\big{\{}\big{[}\bar{n}(\beta_{\text{h}},\omega_{\text{f}})+\frac{1}{2}\big{]}\bar{n}(\beta_{\text{c}},\omega_{\text{i}})+\frac{1}{2}\bar{n}(\beta_{\text{h}},\omega_{\text{f}})+\frac{1}{4}\big{\}},\\
\Pi_{9}^{\text{BM}}&=&\omega_{\text{i}}\omega_{\text{f}}\mathcal{Q}_{\text{e}}\big{\{}\big{[}\bar{n}(\beta_{\text{h}},\omega_{\text{f}})+\frac{1}{2}\big{]}\big{[}\big{(}\bar{n}(\beta_{\text{c}},\omega_{\text{i}})+\frac{1}{2}\big{)}\mathcal{Q}_{\text{c}}-\frac{1}{2}\big{]}+\frac{1}{2}\bar{n}(\beta_{\text{h}},\omega_{\text{f}})+\frac{1}{4}\big{\}},\\
\Pi_{10}^{\text{BM}}&=&\omega_{\text{i}}\omega_{\text{f}}\big{\{}\big{[}\bar{n}(\beta_{\text{h}},\omega_{\text{f}})+\frac{1}{2}\big{]}\bar{n}(\beta_{\text{c}},\omega_{\text{i}})+\frac{1}{2}\bar{n}(\beta_{\text{h}},\omega_{\text{f}})+\frac{1}{4}\big{\}},
\end{eqnarray}
and $\Pi_{1,3,5}$ do not change. With these approximate expressions of $\Pi^{\text{BM}}_{i}$ at hand, we confirm that the exact expression of $\langle W^{2}\rangle-\langle W\rangle^{2}$ reduces to the Born-Markov approximate form in Eq.~(\ref{eq:eqw2bm}).

In the adiabatic limit $\mathcal{Q}_{\ell}=1$, we find the fluctuation reads
\begin{equation}\label{eq:eq83}
\epsilon_{\text{adi}}^{2}\simeq1+\frac{1}{v(\infty)},
\end{equation}
where we have used $v(\infty)\gg\bar{n}(\beta_{\text{c}},\omega_{\text{i}})$. The corresponding Born-Markov approximate result is
\begin{equation}\label{eq:eq84}
\epsilon^{2}_{\text{BM,adi}}\simeq1+\frac{1}{\bar{n}(\beta_{\text{h}},\omega_{\text{f}})},
\end{equation}
where we have assumed $\bar{n}(\beta_{\text{h}},\omega_{\text{f}})\gg\bar{n}(\beta_{\text{c}},\omega_{\text{i}})$. As shown in the main text, we have verified the correctness of Eq. (\ref{eq:eq83}) and Eq. (\ref{eq:eq84}) through numerical calculations. Thus, as long as the condition of $v(\infty)\gg\bar{n}(\beta_{\text{h}},\omega_{\text{f}})$ is satisfied, the non-Markovianity induced noncanonical effect can reduce the fluctuation. Both our numerical simulations and Ref.~\cite{PhysRevE.90.022122} confirm that the above criterion can be well fulfilled, especially in the vicinity of the critical point of bound state. 
The above discussion provides a clear physical picture for our results displayed in the main text.

\end{document}